\documentclass[12pt,preprint]{aastex} 
\usepackage{graphicx} 
\usepackage{natbib}
\usepackage{longtable}

\newcommand{\skipthis}[1]{}

\def\nh3{$\rm{NH_3}$}
\def\NH3{$\rm{NH_3}$} 
 
\def\msun{M$_\odot$}
 
\def\lsun{L$_\odot$} 
\def\kms-1{km~s$^{-1}$}
\def\h2{$\rm{H_2}$} 
\def\CM2{$\rm{cm^{-2}}$}
\def\cm3{$\rm{cm^{-3}}$} 

\def\h2co{H$_2$CO} 
\def\hc3n{HC$_3$N} 

\def\n2hp{N$_2$H$^+$}
\def\n2dp{N$_2$D$^+$}
\def\ch3oh{CH$_3$OH}
\def\c18o{C$^{18}$O}

\newcommand{\lsim}{${\raisebox{-.9ex}{$\stackrel{\textstyle<}{\sim}$}}$ }

\begin{document}

%\slugcomment{\emph{Draft \today}}

\title{ Fragmentation of Molecular Clumps and Formation of Protocluster}

\author{Qizhou Zhang\altaffilmark{1}, Ke Wang\altaffilmark{2}, Xing Lu\altaffilmark{1,3}, Izaskun Jim\'enez-Serra\altaffilmark{2}}

\altaffiltext{1}{Harvard-Smithsonian Center for Astrophysics, 60 Garden Street,
Cambridge MA 02138, USA. E-mail: qzhang@cfa.harvard.edu}

\altaffiltext{2}{European Southern Observatory, Karl-Schwarzschild-Str. 2, D-85748 Garching bei M\"unchen, Germany}

\altaffiltext{3}{School of Astronomy and Space Science, Nanjing University, 22 Hankou Road, Nanjing 210093, China}


\begin{abstract}
Sufficiently massive clumps of molecular gas collapse under self-gravity and fragment to spawn a cluster of stars that have a range of masses. We investigate observationally the early stages of formation of a stellar cluster in a massive filamentary infrared dark cloud, G28.34+0.06 P1, in the 1.3mm continuum and spectral line emission using the ALMA.  Sensitive
continuum data reveal further fragmentation in five dusty cores at a resolution of several 10$^3$ AU. Spectral line emission from \c18o, \ch3oh, $^{13}$CS, \h2co and \n2dp are 
detected for the first time toward these dense cores.
We found that three cores are chemically more evolved as compared with the other two; interestingly though, all of them are associated with collimated outflows as suggested by evidence from the CO, SiO, \ch3oh, \h2co and SO emissions.
The parsec-scale kinematics in \nh3 exhibit velocity gradients along the filament, consistent with accretion flows toward the
clumps and cores. The moderate luminosity and the chemical signatures indicate that the five cores harbor low- to intermediate-mass protostars that likely become massive ones at the end of the accretion. { Despite the fact that the mass limit reached by the $1\sigma$ dust continuum sensitivity} is 30
times lower than the thermal Jeans mass, there is a lack of a distributed low-mass protostellar population in the clump. Our observations indicate that in a protocluster, 
low-mass stars form at a later stage after the birth of more massive protostars.

\end{abstract}

\section{Introduction}

{ Infrared dark clouds (IRDCs) are believed to represent the earliest stage in the formation of massive stars. Observations in different bands of the infrared wavelength have significantly improved our understanding of the dynamics of these clouds.} The early identifications of IRDCs based on the ISO and MSX data \citep{perault1996, egan1998} are extended by observations with the Spitzer Space Telescope. Higher resolution images in the Spitzer IRAC bands reveal intricate details of IRDCs thanks to galactic plane surveys such as GLIMPSE \citep{churchwell2009}. Observations by the Herschel Space Observatory provide a complementary view of IRDCs at far IR wavelengths \citep{molinari2011,wilcock2012a,ragan2012b}. Some of the IRDCs remain in absorption at the 70 $\mu$m band of the Herschel Space Observatory, and transition to emission in the 250, 350 and 500 $\mu$m bands. These 70 $\mu$m-dark IRDCs may represent dense clouds at the earliest evolution stage of star formation.

Among the IRDCs, those with masses of $10^3$ \msun\ or higher at a scale of $\sim$1 pc are the prime candidates to study massive star and cluster formation, { which hereafter we refer to as clumps\footnote{This paper follows the nomenclature used in \citet{zhang2009}, and refers a {\it cloud} as an entity of molecular gas of $10 - 100$ pc, a molecular {\it clump} as an entity of $\lsim$ 1 pc
that forms massive stars with a population of lower mass stars, and {\it dense cores} as an entity of 0.01 to 0.1 pc
that forms one or a group of stars, and {\it condensations} as an entity of $\lsim$ 0.01 pc.}.} 
These clumps contain molecular masses similar to those that harbor massive protostars \citep{beuther2002a, beltran2004, walsh2014} and hypercompact HII regions \citep{zhang1998a,kurtz2002,keto2007}, whose { relatively large} luminosities ($> 10^4$ \lsun), the presence of complex molecules \citep{beuther2007d,rathborne2008,rathborne2011}, and massive and energetic outflows \citep{zhang2001,beuther2002b} all point to active massive star formation. In contrast to clouds with embedded massive star formation, IRDC clumps show few signs of star formation despite their large reservoir of molecular gas at high densities of $>10^4$ \cm3. A survey toward 144 IRDC clumps for H$_2$O masers, a signpost of star formation, found only 14\% of maser occurrence in the sample \citep{wang2006}, a frequency much lower than that in high-mass protostellar objects and HII regions \citep{beuther2002a}. In addition, IRDCs have consistently lower gas temperatures and line widths. Studies in \nh3 found that they have a temperature { on the order of} 15 K, and a line width { on the order of} 2 \kms-1\ averaged over a spatial scale of 1pc \citep{pillai2006b,wang2008,ragan2011,ragan2012a}, lower than those in high-mass protostellar objects (HMPOs) and HII regions \citep{molinari1998, beuther2002a,lu2014}. These characteristics, together with their low luminosities, place massive IRDC clumps at an earlier evolutionary stage than HMPOs and HII regions \citep{rathborne2007}.

High resolution observations of IRDC clumps reveal structures at $\lsim$ 0.1pc scales and provide first glimpses of initial physical and chemical states of cluster star formation. IRDCs tend to have high deuteration fractionation and CO depletion \citep{pillai2007,pillai2012, hernandez2011a, fontani2011, chen2010, zhang2009}, as expected in a low temperature and high density environment \citep{caselli1999, caselli2002a, bergin2007}. { Thermodynamic} analysis found that cores in some clumps are sub-virial, with their gas mass a factor of several greater than the virial mass derived from the line width \citep{pillai2011, li2013, tan2013, lu2014}. Considering that these cores are embedded in dense clumps with substantial external pressure, the imbalance between the gravitational mass and the internal support can be even larger. These analyses do not include magnetic fields, which supply additional support. Indeed, in a recent polarization survey of 14 massive clumps with HMPOs, \citet{zhang2014} found that magnetic fields are important during the formation of dense cores. The plane-of-sky component of magnetic fields derived from statistical analysis is typically 1 - 10 mG \citep{girart2009, tang2009b, girart2013, qiu2013, frau2014, qiu2014}. The magnetic field strength required for a virial balance in these IRDCs cores, on the other hand, is on an order of 0.5 mG. Therefore, IRDC cores can still be in a virial equilibrium for a moderate magnetic field \citep{pillai2015}.

{ Understanding the formation of massive cores is an essential precursor to improving our understanding of the formation of star clusters.
For a typical star-formation efficiency of 10 - 30 \% \citep{lada2010}, and for a cluster having a Salpeter type initial stellar mass function (IMF), one would expect a $10^3$ \msun\ clump to form a cluster of mass between 100 to 300 \msun, with at least 75 stars of mass between 0.5 - 20 \msun.
Therefore, significant fragmentation of molecular gas is required in the clump in order to form a cluster of stars. In such a cluster environment, the majority of stars are at stellar masses of $\lsim$ 1 \msun, which corresponds well to the thermal Jeans mass in the clump with a typical temperature of 15 K and a density of $10^4$ \cm3} \citep{bonnell2002, larson2005, teixeira2007, zhang2009, palau2013}. The question arises when it comes to forming stars around 10 \msun\ or larger. Although hot cores surrounding massive protostars often have masses $10 - 10^2$ \msun\ at a scale of $\lsim$ 0.1pc based on direct high angular resolution observations \citep[e.g.][]{cesaroni1999a, beltran2006a, beuther2005c, zhang2007a, jimenez2012}, they may not represent the state of early fragmentation \citep{longmore2010, rathborne2008}. Observations of massive IRDCs provide direct insights into formation of  massive cores. Indeed, significant fragmentation is observed in massive molecular clumps \citep{zhang2009, swift2009, zhang2011, wang2011, wang2012, wang2014}. These studies find that massive cores are typically 10 to $10^2$ times more massive than the thermal Jeans mass, pointing to a significant support from turbulence as well as perhaps magnetic fields. In addition, cores tend to further fragment, indicating that massive stars form in close groups.

This paper extends the study of a massive IRDC G28.34-0.06 (hereafter G28.34) and presents sensitive ALMA observations of continuum and molecular line emission at the 230 GHz band. The cloud,
at a kinematic distance of $\sim$4.8 kpc, contains several $10^3$ M$_\odot$ of
dense gas along the infrared absorption filament extending 6 pc
in the sky \citep{carey2000,
rathborne2006, pillai2006b, wang2008}. Figure 1 presents an overview of this region from the mid, far infrared to 1.3mm continuum, and gas in \nh3. Two prominent dust continuum clumps, P1 and P2,
are revealed  in the 850 and 450 $\mu$m images obtained from the JCMT \citep{carey2000} . Besides these two massive clumps, additional continuum peaks are revealed in the 1.2mm image obtained from the
IRAM 30m telescope \citep{rathborne2006, rathborne2010}.
Despite the fact that P1 and P2 clumps contain a similar amount of dense gas
within 0.3 pc, the P2 region has a high
gas temperature of 30 K, large \nh3 FWHM line width of 3.3 \kms-1\,
and is associated with
a strong 24 $\mu$m point source with far IR luminosity of $10^3$ \lsun, and with emission of complex organic molecules such as CH$_3$OCH$_3$ and  CH$_3$CHO \citep{vasyunina2014}.
This region is in contrast to P1 which has a gas temperature of 13 K, a relatively
narrow \nh3 line width of 2.7 \kms-1, and an upper limit to the luminosity of
$10^2$ \lsun\ \citep[][hereafter Paper I]{wang2008}, and devoids of a complex molecular chemistry. Furthermore, the gas in P1 appears to be
externally heated with
temperatures decreasing from 20 K in the outside of the cloud
to 13 K inside of the cloud. Likewise, the turbulence measured by the
\nh3 line widths
appears to decrease from pc scales to 0.1 scales.
These observations led \citet{wang2008} to suggest that P1 is at a much
earlier stage of massive star formation compared to P2.

The SMA observations at 1.3mm resolved the G28.34 P1 clump into five cores along
the filament with masses from 22 to 64 \msun\ and an average
projected separation of 0.19 pc \citep[][hereafter Paper II]{zhang2009}. Subarcsecond-resolution observations in 0.88mm reveal further fragmentation in two of the five cores \citep[][hereafter Paper III]{wang2011}.  The masses of the fragments are typically a factor of 10 larger than the Jeans mass derived from the average temperature and density of the medium they are embedded in.   Collimated  outflows are detected in the CO 3-2 emission, emanating from each of the five cores whose temperatures indicate little heating by embedded protostars \citep[][hereafter Paper IV]{wang2012}. Besides CO, no molecular line emission is detected in the 2+2 GHz band of the SMA at 230 and 345 GHz. \citet{zhang2009} find strong CO depletion in the region. These previous studies found that cores in G28.34 P1 clump likely harbors intermediate mass protostars, based on the energetics of molecular outflows. Furthermore, these cores follow turbulent fragmentation, rather than thermal fragmentation.

Our previous observations of G28.34 with the SMA do not have adequate sensitivity and/or dynamic range to reliably detect fragments of thermal Jeans mass (approximately 2 \msun). In addition, the chemical properties in these cores are largely unknown  due to a lack of detection of molecular line emission. Here, we present ALMA observations of G28.34 P1 in Band 6 (230 GHz). The continuum data reach a $1 \sigma$ rms of 75 $\mu$Jy, or 0.065 \msun\ at 1.3mm assuming a dust temperature of 15 K. There is a lack of a population of low-mass protostars with masses down to $\sim 0.4$ \msun\ within the clump. The observations also reveal a plethora of molecules arising from five intermediate-mass cores. This paper focuses on the analysis of physical properties of the region, and potential population of low-mass protostars in the protocluster. The chemical properties of dense cores will be discussed in a future publication. The paper is organized as follows: Section 2 presents the observational setup. Section 3 presents results, and in Section 4, we discuss the implication of the data relevant to cluster formation. Section 5 presents the conclusion of the study.

\section{Observations}

G28.34 P1 was observed with Atacama Large Millimeter/Submillimeter Array (ALMA)  on 2012 November 15 in its extended configuration with a total of 29 12-m antennas in the array (Project ID: 2011.0.00429.S). Including time for calibration, the observing session lasted about 1.5 hours. The actual time on-source is 34 mins. The precipitable water vapor (PWV) was about 1.5 mm, resulting in system temperatures of 80 to 100 K at the observing frequencies. The observations employed the Band 6 (230 GHz) receivers in dual polarization mode. The projected baselines range from 15m to 400m (11.5 - 308 k$\lambda$). The resulting synthesized beam is $0.8''$ when using a robust weighting parameter of 0.5. The FWHM primary beam of the ALMA 12-m antennas is approximately 27$''$ at the observing frequencies.
To maximize continuum sensitivity, we used all 4 spectral windows with a bandwidth of 1.875 GHz in each window. The correlator was set in the FMT mode which provides a uniform channel width of  488.3 kHz (0.62-0.67 \kms-1\ at the observing frequencies).
The frequency coverage of the data range  from 218.07 to 219.95 GHz in SPW0, from 215.62 to 217.49 GHz in SPW1, from 230.18 to 232.06 GHz in SPW2, and from 232.82 to 234.70 GHz in SPW3, respectively.  Quasars J1751+096 and J1830+063 were used for bandpass and time dependent gain calibration, respectively. The flux calibration was achieved by observations of Neptune. The visibility data were calibrated in CASA by the ALMA supporting staff. We construct continuum visibility data using the line free spectral channels for the science target G28.34 P1. The continuum emission was self calibrated in CASA to further improve the dynamical range in the maps. The gain solutions from the self calibration were applied to the spectral line data. We found that the improvement in dynamical range in the continuum is about 20\%. The continuum image reaches a $1\sigma$ rms noise of 0.075 mJy in a synthesized beam of $0''.85 \times 0''.64$. The spectral line sensitivity per 0.7 \kms-1\ channel is 2.5 mJy for SPW0, SPW1 and SPW3. The noise in SPW2 is a factor of 3 higher for unknown reasons. Thus we exclude the SPW2 data for the continuum image.

\section{Results}

\subsection{Continuum Emission}

Figure 2 presents the 1.3mm continuum emission of G28.34 P1. The ALMA image reveals 5 continuum emission peaks along the infrared dark filament. The peaks of the continuum emission coincide with the 1.3mm  SMA continuum sources SMA1 through 5 reported by \citet{zhang2009}. There appears to be another continuum peak $5''$ southwest of SMA5. This dust peak is  located at 10\% of the primary beam response of the ALMA 12-m antenna. With a primary beam corrected peak flux of 4.8 mJy, it is below the $3\sigma$ sensitivity limit of the previous SMA observations \citep{zhang2009}. Since this feature is at 10\% of the primary beam response, clean does not work well. We refrain from discussion of this feature.

The superior sensitivity of the ALMA data reveals additional features in the continuum emission.  SMA1, a single continuum source in the 1.3mm and 0.8 mm SMA data \citep{zhang2009, wang2011}, becomes a  group of 4 continuum peaks. The brightest one coincides with the SMA1 position. The other 3 continuum peaks are at fluxes of 4.0, 2.0, 1.9 mJy, respectively. In addition, spatially extended emission is detected around SMA2, SMA3 and SMA4. Although SMA5 is a single continuum peak in the SMA data,  the ALMA image reveals another continuum peak $2''$ to the east of SMA5. Overall, ALMA image reveals fainter and spatially extended emission. The integrated fluxes in the ALMA continuum image is 228 mJy, which is about 60\% higher than the 
integrated flux of 143 mJy in the SMA 1.3mm image reported in \citet{zhang2009}. 

For convenience of cross referencing, as well as following the convention in the previous papers, we name the 5 groups of sources in the ALMA continuum image Core 1 through Core 5,  corresponding to SMA1 through SMA5 in \citet{zhang2009}, respectively. 
To identify structures in the 1.3mm continuum emission, we { employed} dendrogram algorithm to decompose the continuum emission. However, the decomposition misses some obvious emission features. We then resort to fitting elliptical Gaussian to the continuum image, but using the smallest features identified by dendrogram as a starting point of initial guesses for the Gaussian fitting. We identify a total of 38 components in the simultaneous Gaussian fit. The rms in the residual image is $6 \times 75 \mu$Jy, corresponding to a detection limit of about 0.4 \msun. 
We report the coordinates, flux, the size of the objects in Table 1. The association with previous reported SMA continuum peaks is also listed in the table for cross referencing. The brightest components with peak fluxes $>$ 1.9 mJy~beam$^{-1}$ are also labeled in Figure 2.

\citet{wang2012} derived a gas temperature of 9 to 22K using the higher angular resolution \nh3 (1,1) and (2,2) data.
Adopting the dust opacity law in \citet{hildebrand1983} and a dust
emissivity index of $\beta$ = 1.5, appropriate for
massive dense cores based on multi-wavelength observations at mm and submm
wavelengths \citep{beuther2007b}, we obtained gas masses of the objects
ranging from 0.5 to 16 \msun, assuming a gas-to-dust mass ratio of 100. These parameters are also listed in Table 1. The corresponding H$_2$ densities in these sources range from $10^6$ to 
$10^7$ \cm3.

\subsection{Molecular Line Emission}

Previous arc-second resolution observations detected \nh3 and $^{12}$CO emission in the G28.34-P1 region with the VLA and SMA, respectively  \citep{wang2008, zhang2009, wang2011, wang2012}.
Besides CO, the SMA observations in the 230 GHz band did not detect line emission at a $1 \sigma$ rms noise of 100 mJy~beam$^{-1}$ (at a resolution of $1''.2$ and a spectral resolution of 1.3 \kms-1). The ALMA observations in the same band reveal line emission from 19 molecular species including \c18o, $^{13}$CS, \n2dp, DCN, H$_2$S, SiO 5-4, \h2co, \ch3oh, and SO, at a level of $> 5 \sigma$ with a $1 \sigma$ rms noise of 2.5 mJy~beam$^{-1}$. 

Figure 3 presents the line spectra from the SPW0 spectral window covering 215.7 through 217.5 GHz toward Components 9 and 38, the brightest continuum peaks in Core 2 and Core 5 (see Table 1), respectively. The fluxes are corrected for the ALMA primary beam attenuation for direct comparison. Despite the fact that the two cores have a similar flux in dust continuum, the molecular line emission from the two are very different. The line emission from Core 5 is much fainter than from Core 2. Of particular note is that the emission from \c18o and SO is nearly non detected in Core 5. A lack of \c18o emission is consistent with CO depletion in cold and dense environments at the very early stage of evolution as suggested in \citet{zhang2009}. The fact that we do not detect SO in Core 5 may be related to a lack of injection of ice mantles in the gas phase by thermal desorption, i.e., by an increase of temperature of dust in this core. The chemistry of SO is initiated by the injection of H$_2$S from mantles. H$_2$S is destroyed by H and H$_3$O$^+$ to form S and H$_3$S$^+$, which then react with O and OH to form SO \citep{charnley1997}. A lack of SO emission in Core 5 indicates that the protostar(s) embedded are at an even earlier stage of evolution than those in Core 2. 

Figure 4 presents the integrated emission of \c18o, $^{13}$CS, \n2dp, and H$_2$S lines. The line emission of these molecules is detected mostly near the dust cores. The \c18o emission is strongly detected toward Core 1 through Core 4. There is a faint emission toward Core 5. On the other hand, emissions of $^{13}$CS and H$_2$S are seen toward Core 2 through Core 4. The \n2dp emission is detected in the dense ridge along the filament. However, the emission appears to avoid the dust continuum peaks.

Besides the above mentioned molecules that mainly trace gas around dense cores,
\h2co, \ch3oh and SO are also present towards Cores 1 through 5 (see Figure 4). Strong emission from \h2co, \ch3oh and SO line wings $> 3$ \kms-1 from the cloud systemic velocity is detected outside of dust continuum emission. Their spatial coincidence with the outflows in CO 3-2 \citep{wang2011} and 2-1 (see Section 3.4)  indicates their outflow origin. We will discuss outflows in the region based on the CO 2-1 and SiO 5-4 data in Section. 3.4. Cores 2, 3 and 4 are seen in \c18o, $^{13}$CS, \n2dp, H$_2$S, \h2co, \ch3oh and SO, thus appear to be chemically more active than Cores 1 and 5.

\subsection {Core structure and dynamical state}

At a much higher sensitivity than the previous high resolution studies of \citet{zhang2009} and \citet{wang2011}, the 1.3mm continuum data from ALMA reveal additional fragments. Core 1, which is shown as a single dust peak in the SMA observations, exhibits three additional condensations. These structures are likely the outcome of continuing fragmentation upon the formation of cores along the filament. We investigate the { thermodynamical} properties from the parsec-scale clump to the condensations ($\sim$ 0.01pc). We measure the size of the clump, cores and condensations using a two dimensional Gaussian fit to the dust continuum emission. The fitting reports the integrated flux and the size of the structure. We then measure the line width at the same spatial scales. For the clump, we use the 1.3mm dust continuum data from the IRAM 30m telescope \citep{rathborne2006} and the \nh3 data from the Effelsberg 100m telescope \citep{pillai2006b,wang2008}. For dense cores, we use the SMA 1.3mm continuum \citep{zhang2009}, and the \nh3 data obtained from the VLA \citep{wang2008,wang2012}. At the scale of condensations, no spectral line emission was detected before the ALMA observations. In the ALMA data, a group of molecular lines, e.g., SO, \h2co and  CH$_3$OH, is affected by protostellar outflows. Emission from \c18o, $^{13}$CS, \n2dp and H$_2$S is found toward the dust continuum emission. \n2dp is in general a reliable tracer of cold and dense gas. However, its emission is not present toward all condensations, and does not appear to correlate well with the dust continuum peaks. On the other hand, it appears that the \c18o emission coincides with the dust continuum emission well. We measure the line width using the \c18o data and found a typical FWHM of 1.7 \kms-1. This value is consistent with the line width of \n2dp in the vicinity of Core 3, the strongest emission in the map.

From the measurements described above, we compute the gas mass from the dust continuum emission and the virial mass for the clump (size $\sim$ 1 pc), cores (size $\lsim$ 0.1 pc) and condensations (size $\lsim$ 0.01 pc). The virial mass is computed following
$$M_{vir} = 3k {r \sigma_v^2 \over G}.$$ 
Here $r$ is the radius and $\sigma_v$ is the line-of-sight velocity dispersion, and G the gravitational constant. $k = {5-2a \over 3 - a}$ is a correction factor dependent on the density profile $\rho \propto r^{-a}$ \citep{macLaren1988}. For a constant density, $a = 0$, k= 5/3, the above equation can be rewritten as
 $$M_{vir} = 210 ({r \over pc}) ({\Delta V_{FWHM} \over km~s^{-1}})^2 M_\odot. $$
Here  $\Delta V_{FWHM}$ is the FWHM of the line width along the line of sight.
Table 2 lists gas mass, line width, radius and virial mass of the cloud features identified.

As shown in Table 2, the line width measured from the \nh3 (1,1) line decreases from the clump scale of 0.6pc in diameter to 0.05pc in dense cores, and then increases at scales of 0.017pc at which the condensations are identified. The decrease of line width toward smaller spatial scales in this region is first reported by \citet{wang2008} when analyzing the \nh3 line width obtained from the VLA. They found that the \nh3 line width from the VLA alone is systematically smaller than the line width from the \nh3 spectra from the combined VLA and the Effelsberg 100m telescope. In addition, \citet{wang2008} found that the \nh3 line width is inversely correlated with the \nh3 fluxes. The authors interpret the change of line width as indications of turbulence dissipation from the clump to core scales.

The ALMA observations reveal for the first time spectral line information in dust condensations identified with the SMA at 880 $\mu$m and ALMA at 1.3mm. The line width derived from \c18o in condensations are consistently larger than the \nh3 line width in the cores. Since star formation has already taken place in these condensations, the line width can be broadened by infall and rotation in the envelope \citep[e.g.,][]{zhang1997,zhang1998a,zhang1998b}. The \c18o line width can also be enhanced due to an increase of turbulence thanks to the injection of energy from protostellar outflows (see next Section). 

The virial parameter, $\alpha$, defined as $M_{vir}/M_{gas}$, is smaller than 1 in clump, cores as well as condensations for a constant density distribution. Previous observations of G28.34 found a density profile $\rho \propto r^{-a}$ with the power-law index, a=2 \citep{zhang2009,wang2011} { for detected cores}. It is reasonable to assume a similar density profile in condensations. With a power-law index of 2, the virial masses reported in Table 2 are reduced by a factor of 5/3. Applying this correction to the $\alpha$ values in Table 2 leads to $\alpha$ much smaller than 1. Our observations indicate that molecular gas is not in a virial equilibrium from the clump to condensations.

The virial analysis presented above does not include magnetic fields, which can in turn increase the virial mass. Magnetic fields provide additional pressure in the medium. Their contribution to the virial mass for a critical mass-to-flux ratio of { gravitational collapse} is given by $M_B= {\pi r^2 B \over \sqrt{4 \pi^2 G}}$. { Here $M_B$ is the contribution to the virial mass from magnetic fields}. There are no direct measurements of magnetic fields in G28.34 cores. However, recently studies of dust polarization in regions harboring high-mass protostellar objects found that magnetic fields play an important role in the fragmentation of molecular clumps \citep{zhang2014,koch2014}. Detailed analysis reports the plane-of-sky component of magnetic fields of order of 1-10 mG \citep{girart2009, tang2009b, girart2013, qiu2013, frau2014, qiu2014}, which can increase the virial mass significantly \citep[e.g.][]{frau2014}. Recently, \citet{pillai2015} reported a magnetic field strength of 0.27 mG in an IRDC clump G11.11 using the dust polarization data from JCMT. Assuming the same magnetic field flux in the G28.34 P1 clump, we find the mass for a critical mass-to-flux ratio of $1.6 \times 10^3$ \msun. The virial parameter including the effect of magnetic fields becomes 2 in the clump. We cannot evaluate virial parameters for { the cores detected here} since there is no direct estimate of magnetic fields in IRDC cores. { However, for the sake of argument we assume a magnetic field of 0.27 mG, the same as that reported for IRDC G11.11, the virial parameters derived for cores detected here are still smaller than unity. We therefore suggest, a field strength of 2 mG - 5 mG would be required to make virial parameters close to 1 for these cores.}

\subsection {Molecular outflows}

Molecular outflows are detected { using} several tracers including CO, SiO, \h2co, \ch3oh, and SO. Figure 5 presents CO 2-1 and SiO 5-4 emission in outflows. As shown in Figure 5, high velocity line emission is detected around Cores 1, 2, 3, 4 and 5. There appears to be more than one outflow originating from Cores 1, 2, 3 and 4, consistent with the presence of multiple protostars indicated by multiple continuum peaks. We identify outflow pairs by inspecting the channel maps of the CO, SiO, \h2co, \ch3oh, and SO emission. A total of 10 outflows are identified in the region. Of them, three outflows (Outflows 3a, 3b and 4c) were not seen in the previous study by \citet{wang2011}. Table 3 lists the tracers through which they are identified. The CO emission is the most effective tracer for most outflows. However, outflows 1a and 4b exhibit much stronger emission in SiO, \ch3oh and \h2co emission. SiO and \ch3oh are heavily depleted with low abundances ($10^{-12}$ and $10^{-10}$, respectively) in cold and dense regions \citep{martin1992, jimenez2004}. Their abundances are significantly enhanced in protostellar outflows due to shocks { triggered by the interaction between protostellar wind and the ambient medium}, and release the Si and \ch3oh molecule off the dust grain. The enhanced SiO and \ch3oh are seen in both low-mass and high-mass protostellar outflows \citep[e.g. L1157,][]{zhang1995b,qiu2007}. 

Some of the outflows in the region display remarkable collimation. In particular, outflow 2a associated with Core 2 extends more than 40$''$, or 1 pc in the sky. The full extent is not known since it is beyond the FWHM of the primary beam of the ALMA 12m antenna. The width of the CO outflow is 1$''$.1, yielding a collimation factor, defined as the ratio between the major and the minor axis, of more than 35. This outflow, reaching a velocity $\pm$ 50\kms-1\ from the cloud velocity, also has 
a wide angle component seen in the CO V$_{LSR}$ velocities of 90 - 92 \kms-1. The component has an opening angle of 10$^\circ$. The outflow consists of a chain of CO knots appearing symmetrically with respect to the outflow origin, which is also seen in the CO 3-2 emission with the SMA \citep{wang2011}. The projected spacing of the CO knots are $2''.5$, or $1.2 \times 10^4$ AU. These knots likely correspond to an increase in mass ejection related to episodic accretion, as reported in both low-mass and high-mass protostars \citep[e.g.,][]{qiu2007,qiu2009b}. Assuming a jet velocity of 100 to 500 \kms-1, the time scale associated with the episodic ejection is $5 \times 10^2$ to $10^2$ years. The enhanced accretion/mass ejection can be due to disk instabilities triggered by interactions with a companion.

We derive mass, momentum and energy in molecular outflows identified in the CO following the prescription outlined in \citet{zhang2001,zhang2005a}. We assume that the CO emission is optically thin, and a CO to H$_2$ abundance of $10^{-4}$. We adopt an excitation temperature of 18K following \citet{wang2012}. When an outflow is seen in both CO and SiO, the outflow parameters are derived using the CO data only. For outflows that are seen in SiO only, we use the SiO data to derive outflow parameters assuming an optically thin approximation and a relative [SiO]/[H$_2$] abundance of $9.2 \times 10^{-10}$ \citep{sanhueza2012}. This abundance value of SiO yields consistent outflow masses computed from the SiO and the CO emission. The outflow parameters are given in Table 3.

The mass of the G28.34 outflows, without correcting for the optical depth, ranges from 0.054 to 2.8 \msun. Besides the outflow 1a, all other outflows are more massive than 0.15 \msun. These values are about one order of magnitude greater than the typical mass of outflows powered by low-mass protostars \citep{dunham2014}, and are one order of magnitude lower than the outflow mass associated with massive protostars \citep{zhang2005a}. 

Protostellar outflows are connected to mass ejection during the accretion phase of young stellar objects \citep{shang2007}. High angular resolution observations provide spatially resolved images of massive outflows \citep{su2004,qiu2007,qiu2009b} associated with individual protostars. The inferred accretion rate from massive protostellar outflows is typically 10$^{-4}$ \msun~yr$^{-1}$ \citep{qiu2011}. For G28.34 P1, the accretion rate inferred from the outflow rate is around $10^{-5}$ \msun~yr$^{-1}$ for outflow 2a. In order to reach a star of 10 \msun, it takes $10^6$ years for a constant accretion. This time scale appears to be too long. Therefore, it is likely that the accretion rate increases with time as well as with the protostellar mass. { A variable rate of accretion} has been proposed in the theoretical model of massive star formation \citep[e.g.][]{mckee2002}, and is in agreement with observations that suggest mass infall rates  increase with the protostellar mass \citep{zhang2005b}.

\subsection {Velocity structure along the main filament}

Filaments are dominant structures in the interstellar medium and molecular clouds. Recent galactic-wide surveys reinforce this scenario with findings of spectacular network of filaments in both nearby and more distant molecular clouds \citep{churchwell2009, molinari2010b}. The web-like filaments in massive star forming regions \citep{liu2012, galvan2010, galvan2013, busquet2013} could be part of the hub-filament structure \citep{myers2009b} that transports gas and dust to fuel massive star formation at the center the web.

G28.34 P1 is embedded in a filamentary infrared dark cloud that spans 6pc { on the plane of the sky}. We analyze the large scale \nh3 data obtained with the VLA \citep{wang2008}. Figure 6 presents the peak velocity of the \nh3 (1,1) emission in the region. There are complex velocity structures along the major axis of the filament as well as perpendicular to the filaments (near MM10). Toward clump P1, the line peak velocity varies from 78 to 80 \kms-1. To further examine the
gas velocity, we present in Figure 7 the position velocity plot of the \nh3 (1,1) emission. 
The cut of the pv diagram is along the main axis of the filament with a reference position  $(0'', 10'')$ from Core 3, and a position angle of $44^\circ$. We also compute the centroid velocity using the main component of the \nh3 (1,1) emission, following the formulation of the first moment $v_c = \int F(v) v dv / \int F(v) dv$. Here, $F(v)$ is the flux density, and $v$ is the line-of-sight velocity of the \nh3 (1,1) emission. We used a flux threshold of $5 \sigma$ noise (10 mJy) when computing the moment data to avoid the contamination from the satellite hyperfines. The right panel of Figure 7 plots the line-of-sight velocity along the main axis. We also mark the spatial locations of the dust continuum sources. The filament spans a velocity range from 78 to 82 \kms-1, { which may arise from large scale gas motions in the filament.}

The position velocity diagram and the \nh3 centroid velocity shown in Figure 7 display velocity shifts along the main axis of the filament. There are four mm continuum clumps along the filament, MM4, MM9, MM10 and MM14, reported in \citet{rathborne2010}. MM4, which is G28.34 P1, and MM10, MM14 are at velocities around 79 \kms-1, while MM9 are at velocity around 80 \kms-1. There is a large scale velocity gradient of 0.6 \kms-1 over an angular scale of $80''$ from MM10, MM14 to MM4. In addition, there is a sea-saw velocity pattern with a peak-to-peak amplitude of 0.4 \kms-1. Similarly, the filament near MM9 shows sea-saw velocity patterns with an amplitude of 1 \kms-1. The velocity pattern is consistent with gas flows along the filament toward the clumps and cores \citep[e.g.][]{hacar2011}

\section {Discussion}

\subsection {Fragmentation of massive molecular clumps and formation of massive cores}

High angular resolution observations of massive IRDC clumps { in the recent past} \citep{wang2008,zhang2009,swift2009,rathborne2011,zhang2011,wang2011,wang2012,wang2014,tan2013,sanhueza2013} reveal the physical and chemical state of massive star formation at { an early stage of protostar formation}. Despite typical source distances of kilo-parsecs, interferometric observations achieve sufficient linear resolution to spatially resolve the global thermal Jeans length in molecular clumps (0.15pc for a density of $3 \times 10^4$ \cm3 and a temperature of 15K \citep{pillai2006b,wang2008}). Dense cores revealed in dust continuum emission contain masses at least a factor of 10 larger than the global thermal Jeans mass in the clump\footnote{For cores with masses exceeding the thermal Jeans mass, we refer them as super Jeans cores.} \citep{zhang2009,rathborne2010,sadavoy2010,zhang2011,csengeri2011,wang2011,sanhueza2013, wang2014}, but { comparable} with the turbulent Jeans mass { of the clumps; the sound speed in this instance} is replaced by the turbulent line width. { Observations at higher angular resolution suggest that the detected cores themselves are sub-fragmented.} In IRDC G28.34 and IRDC G11.11, for example, dusty cores first identified at a resolution of 0.05pc were resolved into several continuum peaks { at a spatial resolution on the order of 0.01 pc} \citep{wang2011, wang2014}. These observations indicate that the super Jeans cores continue to fragment and lead to the formation of a group of stars. { However, this need not always be the case. Some super Jeans cores remain starless, but exhibit oscillations in the radial direction\citep[e.g.][]{anathpindika2013}, as demonstrated in hydrodynamic models for such cores \citep[see also,][]{keto2014}.}

A similar trend is also observed when examining the relation between the mass of fragments and the separation. \citet{wang2014} analysed observations of 4 IRDC clumps and found that the fragment mass and separation follow the turbulent Jeans fragmentation if the sound speed in the { Jeans formula is replaced by the turbulent line width. This fragmentation is significantly different to filaments typically observed in low-mass star-forming clouds. Prestellar cores along these filaments are usually separated by a thermal Jeans length \citep[see e.g.][]{arzoumanian2011}.}

These observations offer direct comparison to theoretical models on massive star and cluster
formation. Observations find that massive stars and clusters form in higher density
and more turbulent regions of molecular clouds as compared to their low-mass
counterparts. \citet{bonnell2002} propose in the competitive accretion model
that clouds fragment initially into cores of thermal Jeans mass. These cores subsequently form
low-mass protostars that accrete the distributed gas from a
reservoir of material in the molecular clump. Prototars located near
the center of the gravitational potential accrete at a
higher accretion rate because of a stronger gravitational pull,
thus, experience a faster mass growth. This competitive accretion
model reproduces the stellar IMF observed \citep{bonnell2004}.

Alternatively, \citet{mckee2002} put forward a turbulent core model, in
which stars form via a monolithic collapse of a massive core. In this model, cores are supported by turbulence in a virial equilibrium and have masses much larger than the thermal Jeans mass. Protostellar feedback such as
heating from the embedded protostars increases the gas temperature,
and thus, suppress fragmentation. Therefore, cores harboring massive stars { usually do not sub-fragment} significantly and form one or a few stars \citep{krumholz2005a, krumholz2007}.

{ Observations of typical cores found in IRDCs suggest that relatively massive cores could possibly sub-fragment to form smaller cores which then spawn protostars. The embedded protostars} continue to accrete material from their respective surrounding environment. This picture is similar to the competitive accretion model in that cores do not acquire all their mass before the birth of a protostar, but continue to gain mass during the protostellar accretion. { However}, massive cores contain super Jeans mass, unlike what is assumed in the competitive accretion model, but it is close to the turbulent core model. In the meantime, massive cores appear to fragment and form a group of stars, in contrast with the picture of monolithic collapse. The \nh3 temperature measured in these cores are typically $< 20$ K,
indicating that protostellar heating is ineffective in suppressing fragmentation in the core \citep[see also][]{longmore2010}. While magnetic fields can potentially play an important role in suppressing fragmentation \citep{palau2013, palau2014, zhang2014},
turbulence provides significant support in these cores. The \nh3 data from the VLA observations
\citep{wang2008} demonstrate that despite turbulence decay,
the line widths measured at $3'' - 4''$ scales appear to be large
enough to support the cores in G28.34 P1. Once the protostars are formed, outflows inject energy in to the immediate surrounding of protostars, and increase the line width as seen in the \c18o line width.  This feedback provides additional support that may stop further fragmentation in the gas at the early stage of a cluster formation \citep{wang2010}.

{ One of the key assumptions in the monolithic collapse model is that cluster forming clumps are approximately in hydrostatic equilibrium \citep{mckee2002}. This is based on the observational fact that the time scale for star formation is typically several dynamical times (the free-fall time scale of the gas).} The virial analysis of the G28.34 P1 clump and cores and condensations within { the core} reveals that these entities are far from a virial equilibrium. The virial parameter, defined as $M_{vir}/M_{gas}$, is less than 0.47 from the clump to cores and condensations (See Table 2). This finding suggests that { molecular gas in the clump  is not in a virial balance during star formation}. Similar findings of sub-viral parameters are also reported recently \citep{pillai2011,li2013, tan2013, lu2014}. However, a key component that is not accounted for in the virial analysis is magnetic fields. A moderate field of 0.5 mG or larger can bring the virial parameter to closer to 1. Although magnetic fields have not been measured directly in IRDC cores as present day interferometers lack the necessary sensitivity, such a field strength is reported in more evolved regions such as hot molecular cores  \citep[e.g.][]{zhang2014}. As ALMA begins to offer continuum polarization capabilities now and line polarization in the future, it is expected that direct constraints on magnetic field strengths in regions at the early stages of cluster formation will become available in the next several years.

\subsection {Formation of massive stars through low- to intermediate-mass stages}

G28.34 P1 has a luminosity of $10^2$ \lsun,  a gas temperature of $< 20$ K at a spatial scale of 0.1pc, and a relatively small line width of $<$ 1.7 \kms-1 in FWHM.  In the same complex, G28.34 P2, a molecular clump of 880 \msun\ with a luminosity of $10^3$ \lsun\ and an \nh3 temperature of 45 K, has embedded protostellar object(s) around 8 \msun \citep{zhang2009}. The relatively moderate luminosity, low gas temperature and smaller line widths indicate that G28.34 P1 is at an earlier evolutionary stage than P2 which already has embedded massive protostars. The large reservoir of molecular gas in P1 demonstrates { its potential to form stars in future}. Based on the empirical mass-size relation \citep[e.g.,][]{kauffmann2010a}, G28.34 P1 will likely bear massive stars when accretion is complete.

While the comparison of physical properties between G28.34 P1 and P2 places P1 at an evolutionary stage prior to the presence of a massive protostar, the spectral line data from ALMA further constrains the evolutionary stage of P1. The SMA observations presented in \citet{zhang2009} did not detect molecular line emission besides the CO 2-1 line in the P1 clump at a sensitivity of 100 mJy~beam$^{-1}$ (or 0.4 K), in contrast to the line emission from complex organic molecules in G28 P2 \citep[see also][]{vasyunina2014}. The ALMA observations detect a plethora of spectral lines from 19 molecules at a $1 \sigma$ flux sensitivity of 2.5 mJy~beam$^{-1}$ (0.07K). The line emission includes molecules such as \ch3oh that has a low abundance of $\sim 10^{-10}$ in dense and cold environment thanks to its depletion to dust grains \citep{jimenez2005}. Protostellar heating and/or shocks from outflows release \ch3oh to the gas phase and enhance its chemical abundance. Likewise, SO and OCS abundances are enhanced as a consequence of the release of ice mantles into the gas phase by protostellar heating \citep{jimenez2012}. Therefore, the detection of complex organic and sulfur-bearing  molecules serves as a valuable indicator of protostellar heating.

A detailed chemical modeling of the spectral lines detected in G28.34 P1 is beyond the scope of the present work and best left for a future article. In order to constrain its chemical evolution, we resort to observations of an intermediate-mass protostar ($L \sim 10^2$ \lsun)  in the DR21 filament obtained with the SMA (Zhang et al., in preparation). The SMA data consist of compact and subcompact configurations with a synthesized beam of $3''$. We convolve the SMA data to the same linear resolution as that of the G28.34 spectra in Figure 3, and scale the fluxes according to $(1/D)^2$ to account for the distance difference between the two sources. Here, we adopt a distance of D=1.5 kpc  for DR21 \citep{girart2013}. Figure 3 presents the template spectra of the intermediate-mass protostar in the DR21 filament after the flux scaling.
The template spectra reveal line emission from molecules \c18o, \ch3oh, \h2co, \hc3n and SO, indicative of protostellar heating in the core. The line fluxes in the template spectra match the observed fluxes in Core 2 well, but are stronger than  the line emission in Core 5. This comparison indicates that Cores 2, 3 and 4 harbors intermediate-mass protostars, while Cores 1 and 5 are less active chemically, and may have low-mass protostars embedded or are at an earlier stage of evolution, although this may be inconsistent with the dynamical age of $10^4$ yrs of the outflow in Core 5.

\subsection {Formation of low-mass stars in a cluster}

The clump G28.34 P1, with a mass of $10^3$ \msun\ and an average density of $7 \times 10^4$ \cm3\ can potentially form a cluster of stars with a total mass of 100 to 300 \msun, assuming a 10\% to 30\% star formation efficiency. If the stars follow a Salpeter IMF, one expects 75 to 223 stars in the range of 0.5 to 20 \msun. Among these stars, about 10\% of them could possibly have stellar masses greater than 10 \msun, and 90\% of stars have stellar masses less than 10 \msun, and 60\% of stars have masses between 0.5 and 1 \msun.

The ALMA observations of G28.34 P1 reveal strong dust continuum emission along the ridge of the filament. The spectral line data and the presence of complex organic molecules indicate that  these cores have embedded intermediate-mass protostars. In addition, some of the lower-mass fragments (see Table 1) may harbor low-mass stars. We use the dust continuum data presented in Figure 2 and Table 1 to construct the probability function of the mass distribution. { Figure 8 presents the cumulative mass function ($ N (>M) \sim M$, here $N$ is the number of sources in the mass range $ > M$) for the 38 objects identified.} The dashed line denotes the slope of the Salpeter IMF. As shown in Figure 8, if the cumulative mass function follows the shape of the stellar IMF, then there is a large deficit of lower-mass cores in G28.34 P1. In the mass range of 1-2 \msun\ the number of cores is more than 5 times lower than the expected Salpeter slope.

If low-mass stars in a cluster arise from thermal Jeans mass fragmentation \citep{larson2005, bonnell2002, palau2013},  G28.34 P1 has a global thermal Jeans mass of 2 \msun, and Jeans length of 0.1 pc.
The ALMA continuum image at 1.3mm reaches a $1 \sigma$ rms noise of 0.075 mJy~beam$^{-1}$. For an average temperature of 15 K derived from \nh3 observations \citep{wang2012}, this flux corresponds to a mass of 0.065 \msun, assuming a dust opacity law of \citet{hildebrand1983}, a dust emissivity index of 1.5, and a dust to H$_2$ ratio of $1 : 100$.
The $1 \sigma$ rms noise in the ALMA 1.3mm continuum image is a factor of 30 times lower than the global thermal Jeans mass in the molecular clump. The linear resolution afforded by ALMA observations is 0.02 pc, much smaller than the Jeans length. Therefore, a lack of detection of cores 1-2 \msun\ is significant.

In order to examine the limitation of dynamic range in the ALMA image, we perform simulated observations using the clean model of the 1.3mm continuum image shown in Figure 2, derived from {\it clean} in CASA. To test the ability of recovering low-mass class 0 protostars at the distance of G28.34, we also include a low-mass protocluster NGC 1333 in the model. For NGC 1333, we use the 870 $\mu$m continuum data from the JCMT archive \citep{kirk2006} to derive the sky model at the 1.3mm dust continuum. While the 1.1mm continuum data for NGC 1333 are also available in the literature \citep{enoch2006}, and the data are closer to the frequency of the G28.34 continuum from ALMA, the 870 $\mu$m data are at a resolution of $19''.9$, which offer better spatially resolved structure for the sky model. Figure 9 presents the 870 $\mu$m continuum emission of NGC 1333. Marked in the figure are class 0 protostars identified by \citet{sadavoy2014}. The masses of these cores are 0.5 to 3 \msun.

To construct the sky model from the JCMT data, we first scale the 870 $\mu$m fluxes to 1.3mm by a factor of 3 derived from the comparison of CSO and JCMT data. This scaling factor corresponds to a dust emissivity index of approximately 1.5. In addition, we scale the fluxes of NGC 1333 by $(1/D)^2$ to account for the distance difference between the two sources. From a source distance of 235 pc for NGC 1333 to that of G28.34, the fluxes are reduced by a factor of 417, i.e., $(4800/235)^2$. The flux-scaled image is then deconvolved with a Gaussian beam with a FWHM of $19''.9$ to derive a sky model. The linear scale of $19''.9$ at the distance of NGC 1333 is approximately the linear resolution achieved by the ALMA observations at the distance of G28.34. Therefore, the JCMT data provide adequate source structures that match the ALMA observations.

We perform simulated observations in $casa$ using the sky model. In the simulation, we adopt an array configuration with a total of 29 antennas, a precipitable water vapor PWV = 1.5mm, and an on-source observing time of 34 min, the same parameters during the G28.34 observations. We also change the coordinates of the NGC1333 sky model to the pointing center of G28.34 so that the UV coverage is identical to the G28.34 data.

Figure 9 presents the simulated image using the clean components of G28.34 only, and the one using the clean components plus the sky model of a low-mass protocluster derived from NGC 1333. We reverse the East direction in the RA axis for NGC 1333 to avoid low-mass cores overlapping with the dust emission along the ridge in G28.34. As seen in Figure 9b, the simulated image reproduce the G28.34 image in Figure 2 well. In Figure 9c, the low-mass protostellar cores added to the model are also recovered robustly.

Studies of nearby young clusters \citep{kirk2011} reveal that massive stars are found toward the center of the cluster associated with an enhanced population of low-mass stars in their neighborhood. ALMA observations of G28.34 P1 reveal extended dust continuum along the ridge surrounding the compact sources detected with the SMA at 870 $\mu$m. It is possible that most low-mass protostellar cores form along the dusty ridge. Indeed, the ALMA image at 1.3mm reveals spatially extended emission that is not seen in the SMA image at the same wavelength. The difference in the integrated flux between the ALMA and SMA images is 80 mJy, which is sufficient to form 100 protostellar cores of 0.8 mJy, if assuming 100\% fragmentation.

While the spatially extended emission seen in the ALMA data can give rise to a population of low-mass protostars  in the vicinity of massive stars, a lack of low-mass protostellar cores outside of dense ridge in the P1 region is intriguing.  The absence of a distributed population of low-mass protostellar cores in the clump suggests several possibilities about low-mass star formation in a cluster environment: (1) Low-mass stars only form in the immediate neighborhood of massive stars; (2) The distributed low-mass population does not form in situ, but instead forms outside of the clump, and follows the global collapse and moves to the center of the cluster; and (3) The distributed low-mass protostellar cores have not formed yet in G28.34.

The first possibility can be discounted since it contradicts with the fact that most stellar clusters consist of a population of distributed lower mass stars \citep[e.g.,][]{qiu2007}. The second possibility has been reported in numerical simulations of cluster formation \citep[e.g.][]{smith2013}. In simulations of global collapse of molecular clumps, gas fragments and forms stars along the streams/filaments. These stars follow the gas flow and fall toward the center of the cluster. A similar observational picture was also proposed recently \citep{myers2009b,liu2011,lu2014}, as massive ptotostars are often found near the center of radial gas filaments. In G28.34, the parsec-scale velocity field derived from the \nh3 (1,1) line appears to be consistent with gas flowing toward the center of the cluster. The velocity of the flow, without correcting for the projection effect is $< 0.5$ \kms-1. If the inflowing gas and protostars move at 0.5 \kms-1, it takes about $10^6$ yrs for the low-mass protostars to travel 0.5pc to the central area of the cluster. While we cannot rule out that this process may contribute to the formation of distributed low-mass stars in clusters, its effectiveness in the G28.34 clump is questionable. In particular, within a 2pc area of P1, the higher density gas is mostly distributed within the clump. It becomes counter intuitive that the lower density gas outside of the P1 clump would form stars while the higher density gas within the clump does not. Therefore, we propose that the third possibility is the likely scenario: 
The outer regions of the P1 clump have not fragmented to form low-mass protostellar cores. The limit reached in the ALMA observations is 0.2 \msun\ at a $3 \sigma$ level. A lack of detection of low-mass cores suggests that in clustered star formation, low-mass cores and stars form at a later stage, after the formation of massive stars.

There have been discussions regarding when low-mass stars form relative to massive stars in a cluster \citep{myers2011}. Since massive protostars evolve in a relatively short dynamical time scale of $10^5$ yrs, their strong feedback through radiation and ionization may halt the growth of low-mass stars. With this consideration, low-mass stars should form prior to massive stars. Our observations of IRDC G28.34 P1 and other IRDC clumps indicate that massive stars undergo evolutionary stages of low- to intermediate-mass protostars, which extends their formation time scale. In addition, it appears that massive stars form first in a cluster. Regimes that reach high enough densities become gravitational unstable and proceed to star formation. The high density regimes are normally reached first toward the center of the clump. That explains why massive stars form at the center of a cluster. Thanks to a longer accretion time, protostars who have lower mass initially gain mass over time and become massive stars. Formation of low-mass stars in the clump follows as gas continues to condense while turbulence dissipates in the clump to reach a super critical state. Protostars formed early on in the clump benefit from a longer accretion, thus become massive stars. The accretion rates inferred from molecular outflows are typically $10^{-5}$ \msun~yr$^{-1}$. Should the rate be a constant, it takes more than $10^6$ years to form a 10 \msun\ star. Therefore, it is likely that mass accretion increases over time or accretion could be episodic\citep[e.g.][]{stamatellos2012}.

\section {Conclusion}

We present ALMA observations of a massive IRDC clump G28.34 P1. The clump has a mass of $10^3$ \msun\ at a spatial scale of 0.6pc in diameter, embedded in a long filament stretching over 6pc in the sky. We analyse the ALMA data in conjunction with the (sub)mm continuum data from the SMA and the \nh3 data from the VLA to assess physical properties from the filament to the parsec-scale clump, to the 0.1pc dense cores and to the 0.01pc scale condensations. The main findings are:

(1) The IRDC filament exhibits a velocity pattern consistent with mass accretion along the filament toward clumps and dense cores.

(2) The 1.3mm continuum data from ALMA reveal 5 cores consistent with previous SMA observations. These cores are at least a factor of 10 more massive than the thermal Jeans mass, implying that turbulence and perhaps magnetic fields are important in supporting massive cores during the fragmentation.

(3) For the first time, the ALMA data reveal spectral line emission from 19 molecules including $^{12}$CO, \c18o, \ch3oh, $^{13}$CS, \h2co, \n2dp and SO in the dense cores. Comparison with spectral line data from nearby protostellar cores with embedded intermediate-mass stars indicates that G28.34 P1 undergoes active massive star formation currently at an intermediate-mass stage.

(4) The superior flux sensitivity in the ALMA continuum data reveals additional fragments with masses of a few \msun. Despite a $1\sigma$ mass sensitivity of 0.065 \msun, there is a lack of a wide spread low-mass protostellar population expected from thermal Jeans fragmentation. This finding indicates that low-mass protostars form after massive protostars in a protocluster.

(5) { Cores detected in this cloud have masses at least an order of  magnitude larger than the thermal Jeans mass and appear to be mostly supported by a turbulent velocity field.}
This study indicates a successive star formation in a protocluster, with cores harboring massive stars form first as gas in these cores becomes super critical when gravity overcomes the internal support. Star formation spreads as gas in other parts of the clump become super critical and collapse to form stars.

\acknowledgements
We thank P. C. Myers for enlightening discussions. This paper makes use of the ALMA data. ALMA is a partnership of ESO (representing its member states), NFS (USA) and NINS (Japan), together
with NRC (Canada) and NSC and ASIAA (Taiwan), in cooperation with the Republic of Chile. The Joint ALMA Observatory is operated by ESO, AUI/NRAO and NAOJ.  This research is partly supported by the National Science Foundation of China under Grant 11328301. K. W. acknowledges support from the ESO fellowship. X. L. acknowledges the support of Smithsonian Predoctoral Fellowship. I. J.-S. acknowledges funding received from the People Programme (Marie Curie Actions) of the European Union's Seventh Framework Programme (FP7/2007-2013) under REA grant agreement PIIF-GA-2011-301538. This study used observations made with the Spitzer Space Telescope, which is operated by the Jet Propulsion Laboratory, California Institute of Technology under a contract with NASA.

\clearpage

\renewcommand{\thefootnote}{\alph{footnote}}
\begin{longtable}{rccccccccc}
\caption{Physical Parameters of Dense Gas Structure} \\
%\begin{center}
\hline \hline
         Source &       RA     &      DEC       &  T & $S_{Peak}^a$ & $S_{int}^a$ &  Mass &   size$^b$  &    PA  & Reference$^c$ \\
                &     (h:m:s)  &   (d:m:s)      &   K    & mJy/bm &   mJy &   \msun\ &   $'' \times ''$ &   ($^\circ$)  \\ \hline  \endhead
    1   &     18:42:51.21  &   -04:03:05.2 &  22.2  &  1.9  &  4.2  &  2.6  &  1.2 $\times$ 0.3  &  137 &  \\
    2   &     18:42:51.13  &   -04:03:06.06  &  18.5  &  2.0  &  10.5  &  7.7  &  1.8  $\times$  1.2  &  60 & \\
    3  &     18:42:51.26  &   -04:03:06.5  &  19.5  &  4.0  &  5.4  &  3.7  &  0.5  $\times$  0.4  &  76  & \\
     4  &     18:42:51.19  &   -04:03:07.3  &  17.9  &  9.3  &  16  &  12.1  &  0.7  $\times$  0.6  &  87 & SMA1 \\ \hline
    5  &     18:42:50.96  &   -04:03:10.1  &  22.8  &  1.6  &  16.6  &  9.9  &  3.4  $\times$  1.4  &  42  & SMA2d \\
    6  &     18:42:50.84  &   -04:03:10.5  &  18.6  &  1.7  &  1.7  &  1.2  &    $-$ &  $-$ &  \\
    7  &     18:42:50.92  &   -04:03:11.3  &  24.3  &  0.9  &  0.9  &  0.5  &  $-$ &  $-$  &  \\
    8  &     18:42:51.20  &   -04:03:11.3  &  12.3  &  0.7  &  1.4  &  1.6  &  1.4  $\times$ 0.3  &  93   & \\
     9  &     18:42:50.86  &   -04:03:11.5  &  20.7  &  7.6  &  9.8  &  6.4  &  0.5  $\times$  0.1  &  20 & SMA2a \\
     10  &     18:42:50.77  &   -04:03:11.6  &  16.8  &  5.3  &  20.1  &  16.4  &  1.5  $\times$  1.0  &  49 & SMA2b \\
    11  &     18:42:50.81  &   -04:03:12.5  &  18.2  &  2.2  &  2.2  &  1.7  &  $-$ &  $-$  & SMA2c \\
    12  &     18:42:50.93  &   -04:03:12.9  &  22.7  &  0.8  &  6.3  &  3.8  &  2.4  $\times$ 1.6  &  113  &  \\
    13  &     18:42:50.70  &   -04:03:12.9  &  16.6  &  1.2  &  1.2  &  1.0  &  $-$ &  $-$  &   \\
    14  &     18:42:50.75  &   -04:03:13.3  &  16.7  &  1.1  &  4.5  &  3.6  &  1.6  $\times$  1.0  &  114   & \\
    15  &     18:42:50.73  &   -04:03:13.8  &  17.4  &  0.8  &  1.4  &  1.1  &  0.9  $\times$  0.3  &  123  &  \\
    16  &     18:42:50.63  &   -04:03:13.9  &  15.7  &  0.7  &  1.2  &  1.0  &  0.7  $\times$ 0.6  &  23  &  \\ \hline
    17  &     18:42:50.69  &   -04:03:15.4  &  16.2  &  2.6  &  16  &  13.4  &  1.9  $\times$  1.5  &  95   & \\
    18  &     18:42:50.31  &   -04:03:15.5  &  19.1  &  1.0  &  1.0 &  0.7  &  $-$ &  $-$   & \\
    19  &     18:42:50.93  &   -04:03:16.0  &  17.5  &  0.4  &  1.3  &  1.0  &  1.4  $\times$  0.7  &  50   & \\
     20  &     18:42:50.58  &   -04:03:16.3  &  18.3  &  8.9  &  13.8  &  10.2  &  0.7  $\times$  0.5  &  71  &  SMA3 \\
     21  &     18:42:50.61  &   -04:03:17.0  &  18.3  &  2.8  &  10.1  &  7.5  &  1.7  $\times$  0.7  &  29  &  \\
    22  &     18:42:50.74  &   -04:03:17.2  &  15.6  &  1.1  &  2.9  &  2.5  &  1.5  $\times$  0.5  &  73  &  \\
    23  &     18:42:50.51  &   -04:03:17.2  &  17.2  &  1.2  &  19  &  15  &  3.3  $\times$  2.3  &  169  &  \\
    24  &     18:42:50.70  &   -04:03:18.2  &  14.9  &  0.8  &  1.5  &  1.3  &  0.9  $\times$ 0.4  &  72  &  \\
    25  &     18:42:50.53  &   -04:03:18.2  &  17.1  &  0.9  &  0.9  &  0.7  &  $-$ &  $-$  &  \\
    26  &     18:42:50.80  &   -04:03:18.7  &  12.5  &  1.3  &  4.5  &  4.9  &  1.4  $\times$  1.0  &  125   & \\
    27  &     18:42:50.20  &   -04:03:19.1  &  17.8  &  0.6  &  1.7  &  1.3  &  2.2 $\times$  0.3  &  99  &  \\ \hline
  \pagebreak
     28  &     18:42:50.28  &   -04:03:20.2  &  18.9  &  5.0  &  5.0  &  3.6  &  $-$ &  $-$  &  SMA4a \\ 
     29  &     18:42:50.23  &   -04:03:20.3  &  18.9  &  6.5  &  19.4  &  14  &  1.4 $\times$  0.7  &  87  & SMA4b \\
     30  &     18:42:50.32  &   -04:03:20.9  &  18.5  &  3.8  &  21.5  &  15.8  &  2.1  $\times$  1.2  &  12  & SMA4c \\
     31  &     18:42:50.39  &   -04:03:21.0  &  18.7  &  3.5  &  10.6  &  7.7  &  1.5  $\times$  0.7  &  113  &  \\
    32  &     18:42:50.39  &   -04:03:22.5  &  17.1  &  1.7  &  10.7  &  8.5  &  1.8  $\times$  1.6  &  6  &  \\
    33  &     18:42:50.49  &   -04:03:22.9  &  16.9  &  0.8  &  1.4  &  1.1  &  1.0  $\times$  0.1  &  173  &  \\
    34  &     18:42:50.73  &   -04:03:22.9  &  11.4  &  0.5  &  0.8  &  0.9  &  0.9  $\times$  0.3  &  92   & \\
    35  &     18:42:50.33  &   -04:03:23.1  &  15.3  &  0.6  &  0.6  &  0.5  &  $-$ &  $-$   & \\
    36  &     18:42:50.30  &   -04:03:23.8  &  14.8  &  0.6  &  3.6  &  3.3  &  2.0 $\times$  1.2  &  134  &  \\ \hline
    37  &     18:42:49.91  &   -04:03:25.2  &  11.6  &  3.1  &  10.9  &  12.8  &  1.7  $\times$  0.8  &  98  &  \\
     38  &     18:42:49.82  &   -04:03:25.5  &  9.2   &  13.9  &  18.2  &  27  &  0.4  $\times$  0.4  &  37  & SMA5 \\
%    component34  &     18:42:49.38  &   -04:03:29.710  &  15.0  &  5.7  &  27.6  &  25.1  &  1.9  $\times$  1  &  135   &      \\ 
\hline
\footnotetext[1]{The fluxes reported here are primary beam corrected. The beam size is $0''.85 \times 0''.64$ at a positional angle of $89^\circ.82$.}
\footnotetext[2]{$-$ represents sources unresolved.}
\footnotetext[3]{Source names used in \citet{zhang2009, wang2011}}
%\end{center}
\end{longtable}

\clearpage

\begin{table}[h]
\caption{Virial  Parameters in the Dense Gas}
\begin{center}
\begin{tabular}{lrrrrrrr} \hline \hline
  Name  & $M_{gas}$  & $\Delta V^a$   &   r  &  M$_{vir}$  & $\alpha^b$ &  M$_{B}$  & $\alpha_{total}^c$ \\
   &  (\msun)  & (\kms-1) & (pc)  & (\msun) &  & (\msun) &  \\ \hline
Clump G28-P1     &       1000.0   &   2.67  &    0.30   &    444   &   0.44   &    1637.9    &   2.08 \\ \hline
Core 1     &      28.0  &     1.2  &   0.030  &     6.93  &    0.25   &    9.79  &    0.60 \\
Core 2     &      21.0   &    1.5  &   0.021   &    9.91 &     0.47  &     8.199  &    0.86 \\
Core 3     &      22.0  &    0.94  &   0.023   &    4.28   &   0.19  &     9.90   &   0.64 \\
Core 4     &      43.00  &    1.10  &   0.028   &   7.07    &  0.16   &    14.44   &   0.50 \\
Core 5      &      20.0   &    1.70 &    0.01  &     6.34  &    0.32 &      2.03  &    0.42 \\ \hline
Condensation 4  &  12.1  &    1.70 &  0.0075   &    4.57   &   0.38   &    1.06    &  0.47 \\
Condensation 9  &   6.4  &     1.70 &   0.0026  &     1.58  &    0.253  &    0.126  &    0.27 \\
Condensation 20  &  10.2  &     1.70 &   0.0069  &     4.17  &    0.41 &    0.880  &    0.50 \\
Condensation 28  &  3.6   &   1.70  &  0.0012   &   0.705    &  0.20   &  0.0251  &    0.20 \\
Condensation 38  &  27.0  &     1.70 &  0.0047  &     2.82  &    0.10  &    0.402  &    0.12 \\ \hline
\end{tabular}
\tablenotetext{a}{The line width for cores is measured from the \nh3 (1,1) data observed from the VLA \citep{wang2012}. Line widths in condensations are measured from the C$^{18}$O 2-1 data in this paper.}
\tablenotetext{b}{$\alpha = {M_{vir} \over M_{gas}}$.}
\tablenotetext{c}{$\alpha_{total} = {M_{vir} + M_B \over M_{gas}}$, where $M_B$ is the magnetic virial mass.}
\end{center}
\end{table}

\begin{table}[h]
\caption{Outflows and their Physical Parameters$^a$}
\begin{center}
\begin{tabular}{llcrrrrr} \hline \hline
 Name     & Tracers & Mass    & Momentum  & Energy & L  &  $t_{dyn}^c$  &  $M_{out}$ (10$^{-5}$) \\
          &         & (\msun) & (\msun \kms-1) & \msun (\kms-1)$^{2}$ & pc  & 10$^3$ yrs & \msun~yr$^{-1}$ \\ \hline
1a$^b$   & SiO & 0.054  & 0.28  &  1.1   & 2.2    & 6.8-6.1   & 0.78  \\ 	 % 
1b   & CO,SiO & 0.65 & 6.7 & 38   & 0.49   &  16-27  & 4.0  \\	\hline % 
2a   & CO,SiO & 3.8  & 75  & 830  & 1.0    &  11-14  &  32 \\ 	 % 
2b   & CO,SiO & 0.15  & 2.3  & 22  & 0.17   &  5.6  & 2.6  \\ \hline	 % 
3a   & CO,SiO & 0.15  & 0.73 &  2.3   & 0.16    & 13   &  1.1 \\ 	 % 
3b   & CO,SiO & 2.1  & 29  &  270   & 0.90    & 4.5-9.5   & 16  \\ 	 %
3c   & CO,SiO & 1.2  & 23  &  253   & 0.74    & 9.2-32   & 12  \\ \hline	 %
4a   & CO,SiO & 0.56  & 8.3  & 77    & 0.51    & 7.0-10   & 7.1  \\ 	 % 
4b$^b$   & SiO &  0.23 &  1.4 &  5.5   & 2.3    & 4.1-12   & 2.2  \\ \hline	 % 
5a   & CO,SiO & 0.30  & 1.9  & 6.4    & 0.40    & 41   & 0.73  \\ \hline	 % 
\end{tabular}
\tablenotetext{a}{Parameters are obtained using CO 2-1 data unless stated otherwise. Data are not corrected for the inclination angle of outflows.}
\tablenotetext{b}{Assuming an SiO relative abundance of $9.2 \times 10^{-10}$ \citep{sanhueza2012}.}
\tablenotetext{c}{If both blue- and red-lobes are present, both dynamical time scales are listed.}
\end{center}
\end{table}

\clearpage

\begin{figure}[h]
\figurenum{1a}
\includegraphics{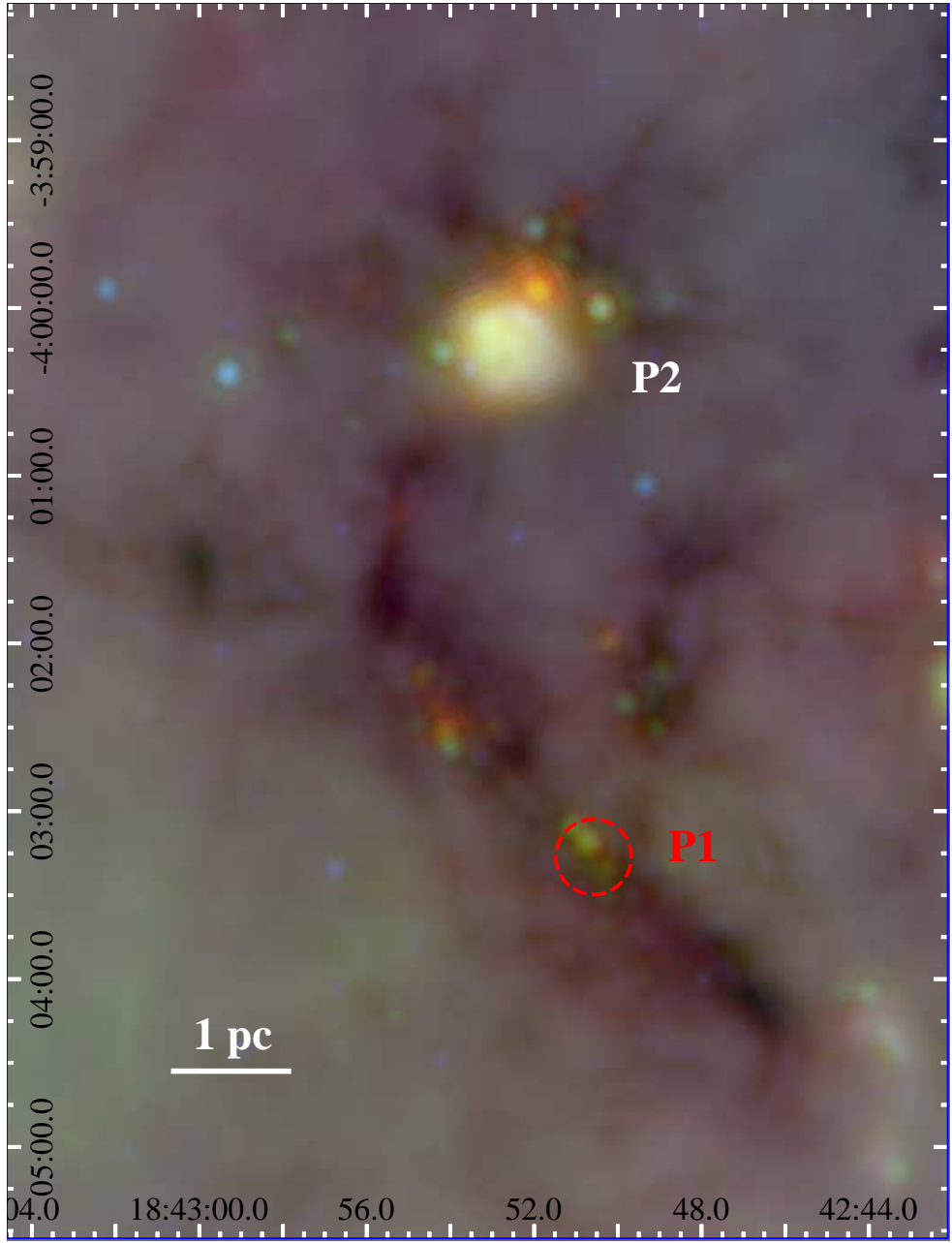}
\caption{Three color Spitzer composite image
(red/green/blue = 70/24/8 $\mu$m) showing the IRDC G28.34+0.06. The circle close P1 marks the FWHM field of view observed with ALMA.}
\end{figure}

\clearpage

\begin{figure}
\figurenum{1b}
\vskip 5in
\includegraphics{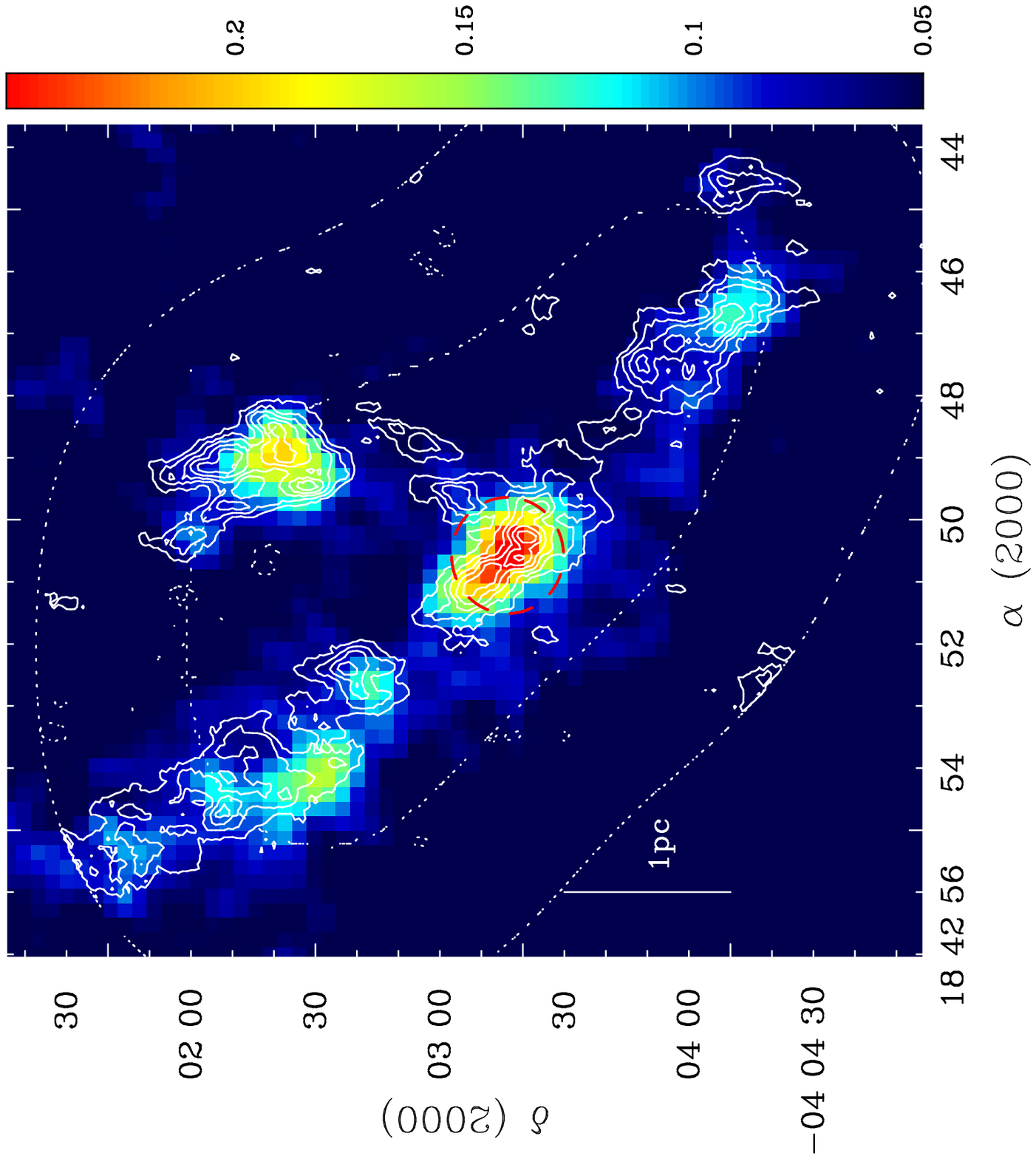}
\caption{
The integrated intensity of the \nh3 (1,1)
emission \citep{wang2008} in white solid contours overlaid on the 1.3mm continuum  \citep{rathborne2010} in color scale. The \nh3 
image is contoured at
10\% of the peak (1 Jy~beam$^{-1}$ $\times$ km s$^{-1}$). The thin dotted lines indicates the
50\% and 100\% of the sensitivity level of the 7 pointing mosaic in \nh3 from the VLA.
The \nh3 data have a resolution of $5'' \times 3''$. The thick dashed
circle marks the P1 field that ALMA observed in this work. The color bar on the right-hand side of the plot denotes the 1.3mm continuum flux scales in Jy~beam$^{-1}$.
}
\end{figure}

\clearpage
\begin{figure}
\figurenum{2}
\vskip 6in
\includegraphics{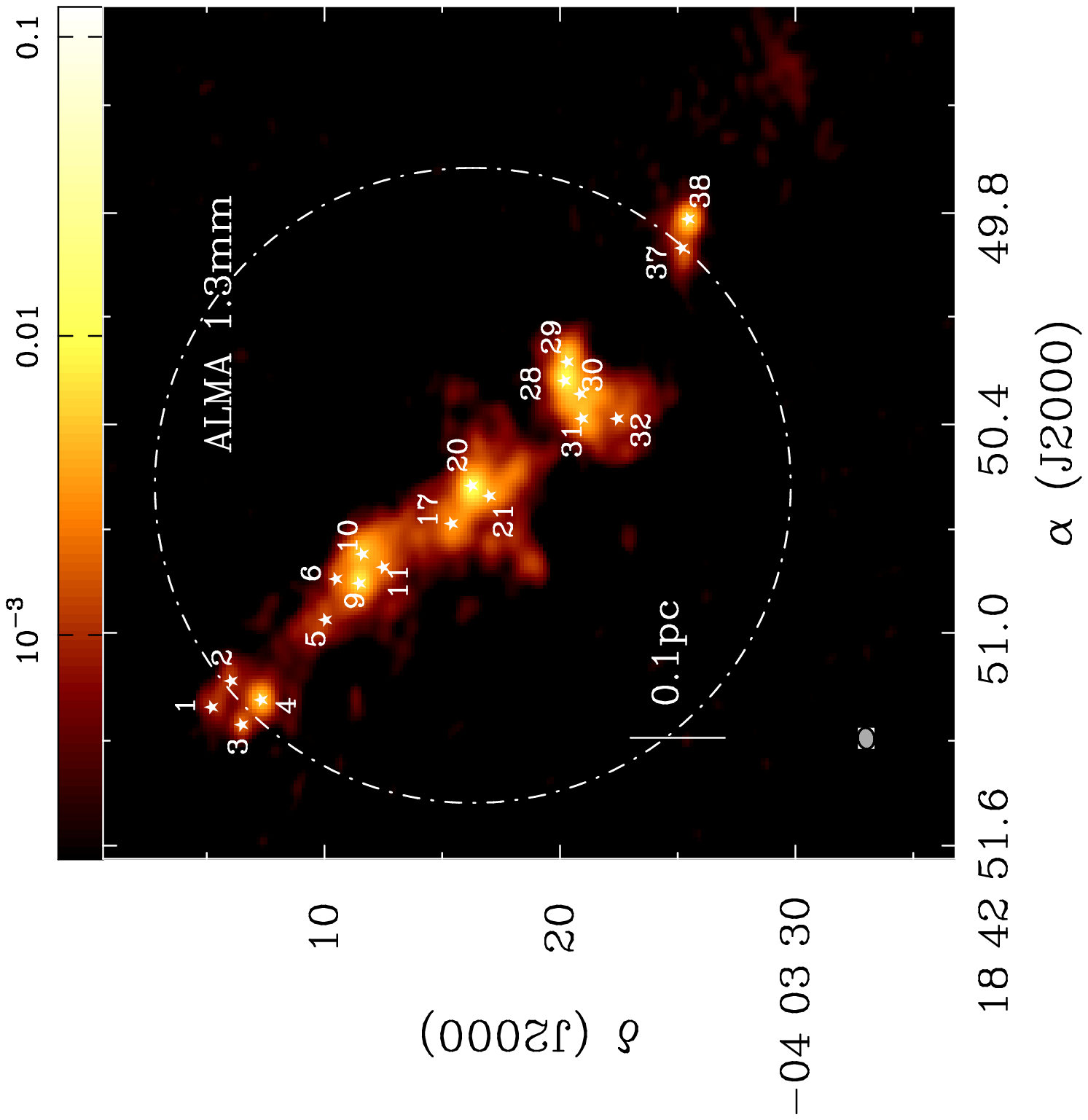}
\caption{1.3 mm Continuum image of IRDC G28.34 P1 obtained with ALMA. Data are plotted in logarithmic scale. The dashed-dotted circle outlines the FWHM of the primary beam of the ALMA 12m antenna. The shaded ellipse at the lower left corner of the panel marks the synthesized beam. The star symbols and numbers mark the brightest dust continuum components listed in Table 1. The wedge at the top of the panel displays the color scales corresponding to fluxes (Jy~beam$^{-1}$)  in the logarithmic scale. Features toward the lower right of the image are at the 10\% response of the primary beam, thus are not robust due to inadequate clean.}
\end{figure}

\clearpage

\begin{figure}[h]
\figurenum{3}
\includegraphics{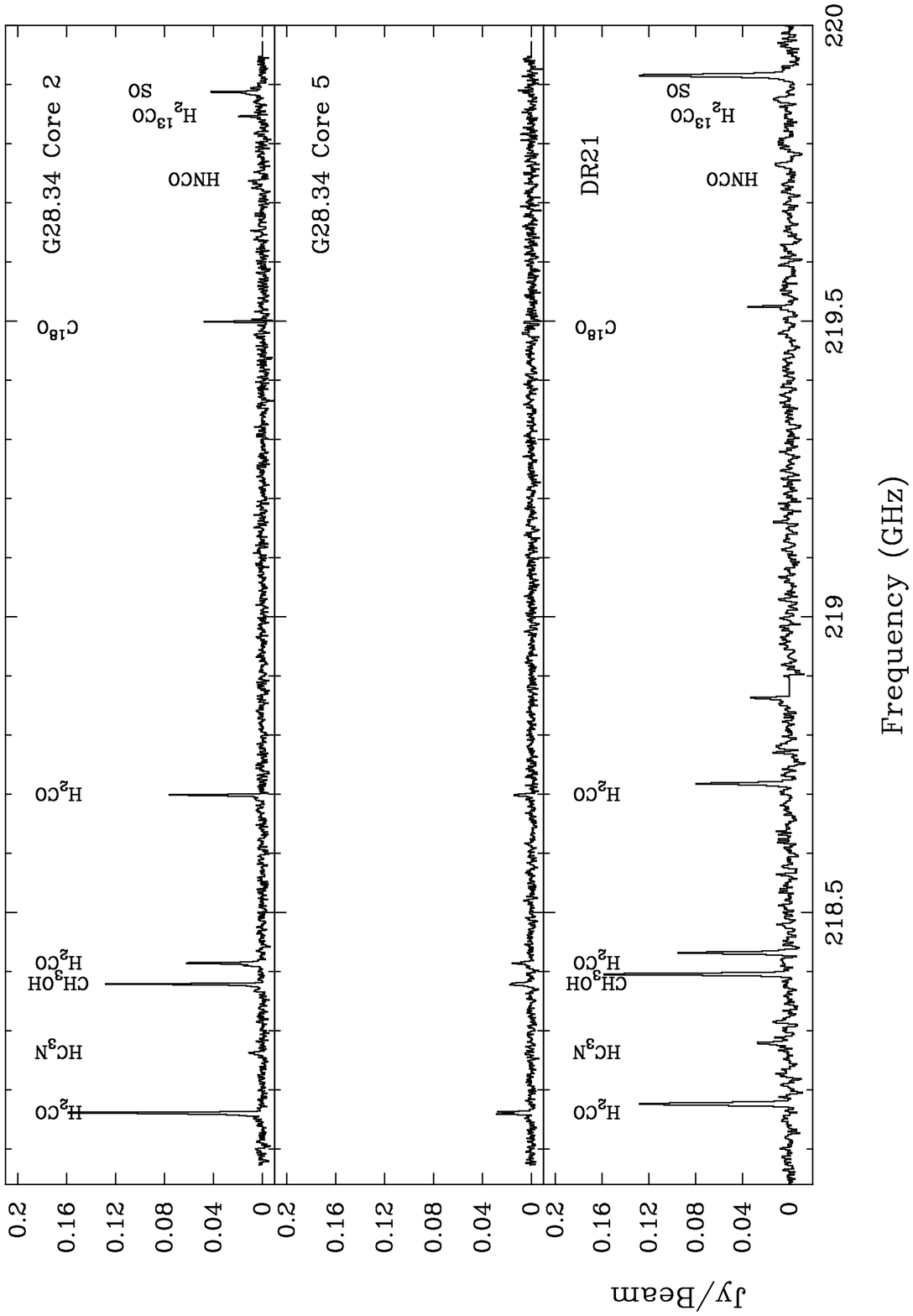}
\caption{ALMA spectra from Cores 2 and 5 (Components 9 and 38 in Table 1, respectively) in G28.34. The flux scale is corrected for primary beam attenuation. Also shown are spectra for an intermediate-mass protostar in the DR 21 filament obtained from the SMA, scaled to the distance of G28.34 for comparison. }
\end{figure}

\clearpage

\begin{figure}[h]
\figurenum{4a}
\includegraphics{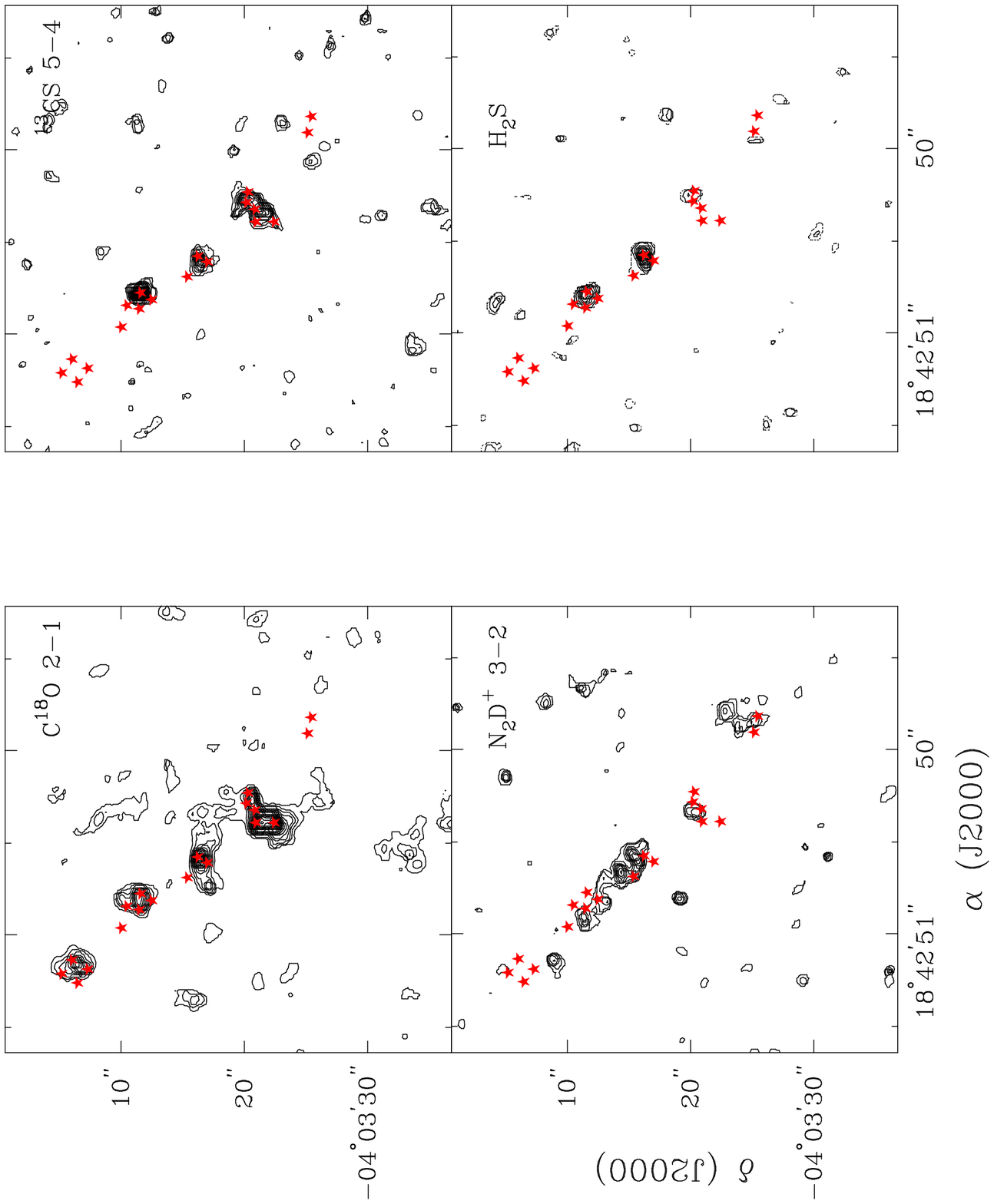}
\caption{Velocity integrated emission (Moment 0) of molecular lines from C$^{18}$O, $^{13}$CS, \n2dp and H$_2$S. The range of integration covers the entire velocity range of the line emission. The contours for C$^{18}$O, $^{13}$CS and \n2dp emission are plotted in equal increment of 10 mJy~\kms-1\ starting from 10 mJy~\kms-1. The contours for H$_2$S emission are plotted in equal increment of 6 mJy~\kms-1\ starting from 6 mJy~\kms-1. The star symbols denote the brightest dust continuum peaks in Table 1. }
\end{figure}
\clearpage

\begin{figure}[h]
\figurenum{4b}
\includegraphics{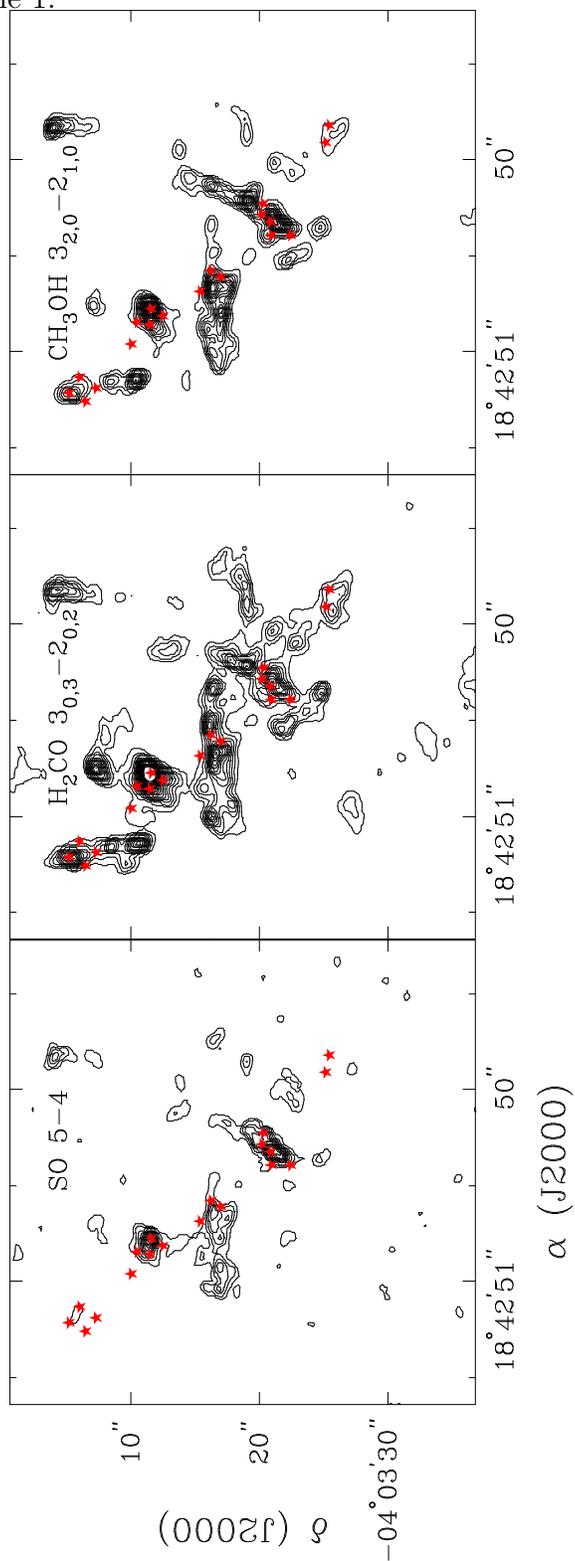}
\caption{Velocity integrated emission (Moment 0) of molecular lines from SO, H$_2$CO, and CH$_3$OH. The range of integration covers the entire velocity range of the line emission. The contours for SO emission are plotting in equal increment of 20 mJy~\kms-1\ starting from 20 mJy~\kms-1. The contours for H$_2$CO, and CH$_3$OH emission are plotting in equal increment of 30 mJy~\kms-1\ starting from 30 mJy~\kms-1. The star symbols denote the brightest dust continuum peaks in Table 1.}
\end{figure}

\clearpage
\begin{figure}[h!]
\figurenum{5a}
\includegraphics{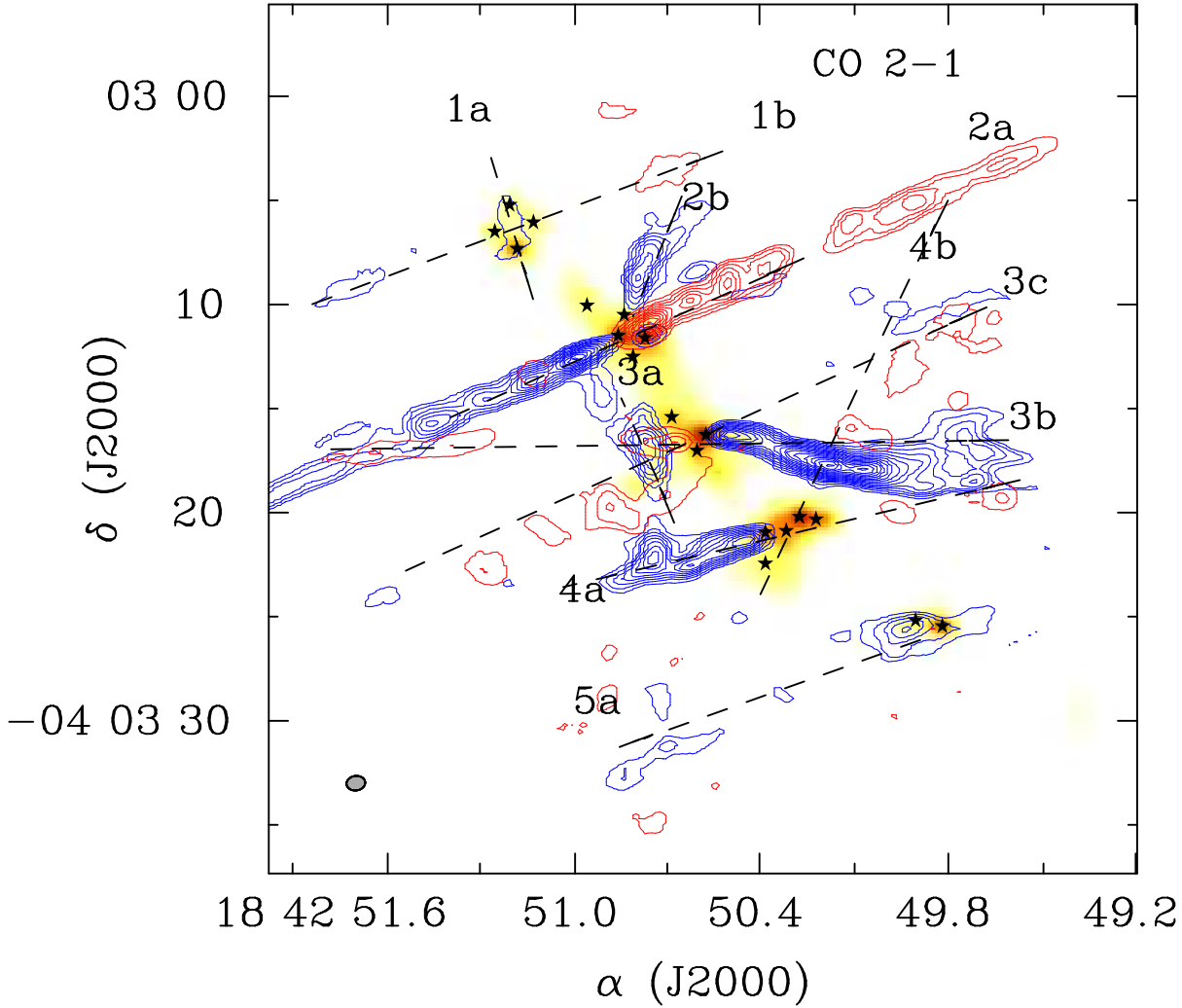}
\caption{Molecular outflows detected in CO 2-1 and SiO 5-4 in the G28.34 P1 region. The blue and red-shifted outflows are plotted in blue and red contours, respectively. The data shown in color scales are 1.3mm continuum emission also shown in Figure 2. The black dashed lines mark the outflows identified in the region. The CO emission is integrated over a velocity range from 54 to 72 \kms-1 for the blue-shifted emission, and 96 to 112 \kms-1 for the red-shifted emission. The SiO emission is integrated over the velocity range from 65 to 76 \kms-1 for the blue-shifted emission,and 82 to 96 \kms-1 for the red-shifted emission.
The contours for the CO emission are plotting in equal increment of 0.5 Jy~\kms-1\ starting from 0.5 Jy~\kms-1. The contours for the SiO emission are plotting in equal increment of 0.02 Jy~\kms-1\ starting from 0.02 Jy~\kms-1. The star symbols denote the brightest dust continuum peaks in Table 1.}
\end{figure}

\clearpage
\begin{figure}[h!]
\figurenum{5b}
\includegraphics{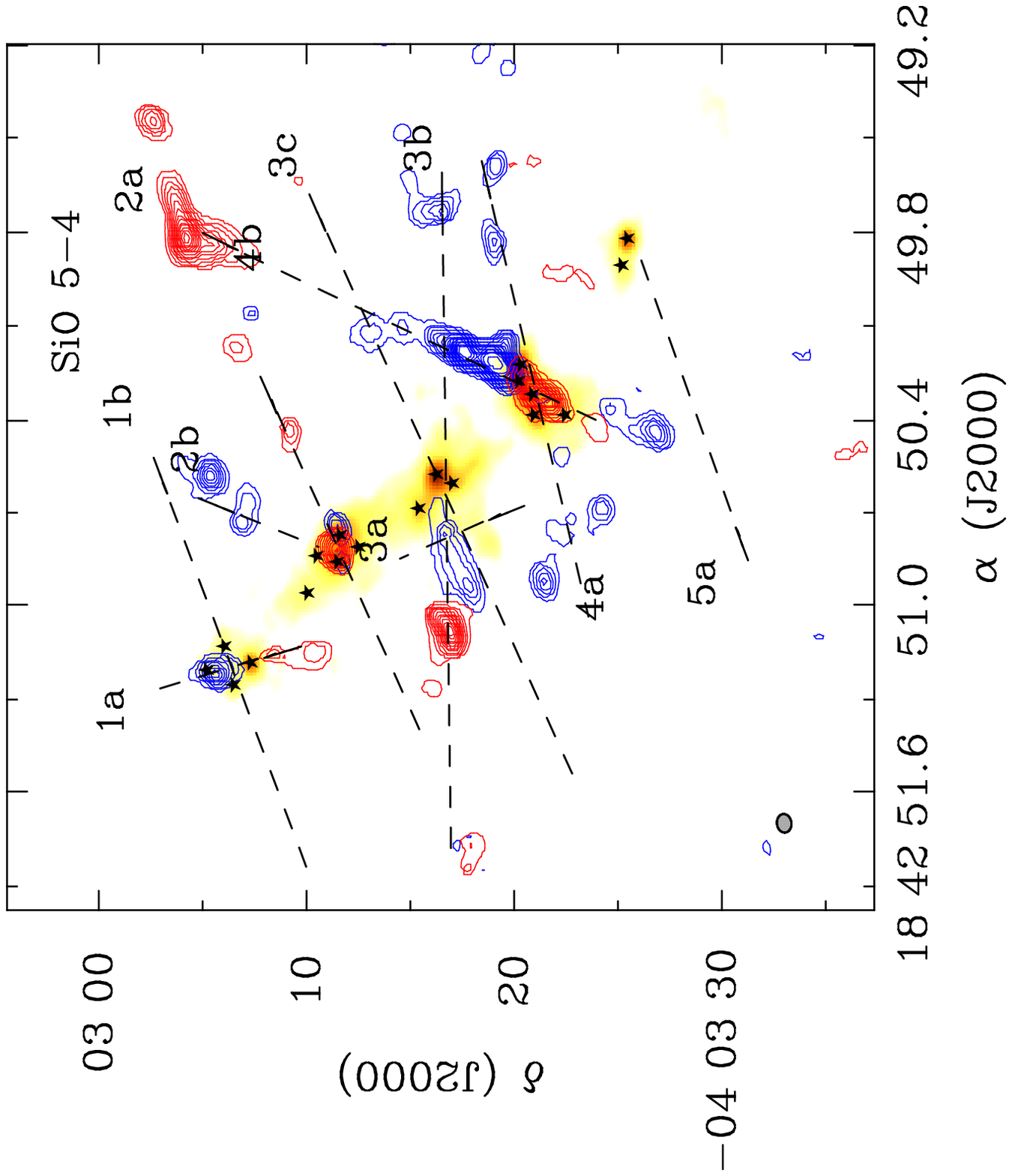}
\caption{}
\end{figure}
\clearpage

\begin{figure}[h]
\figurenum{6}
\includegraphics{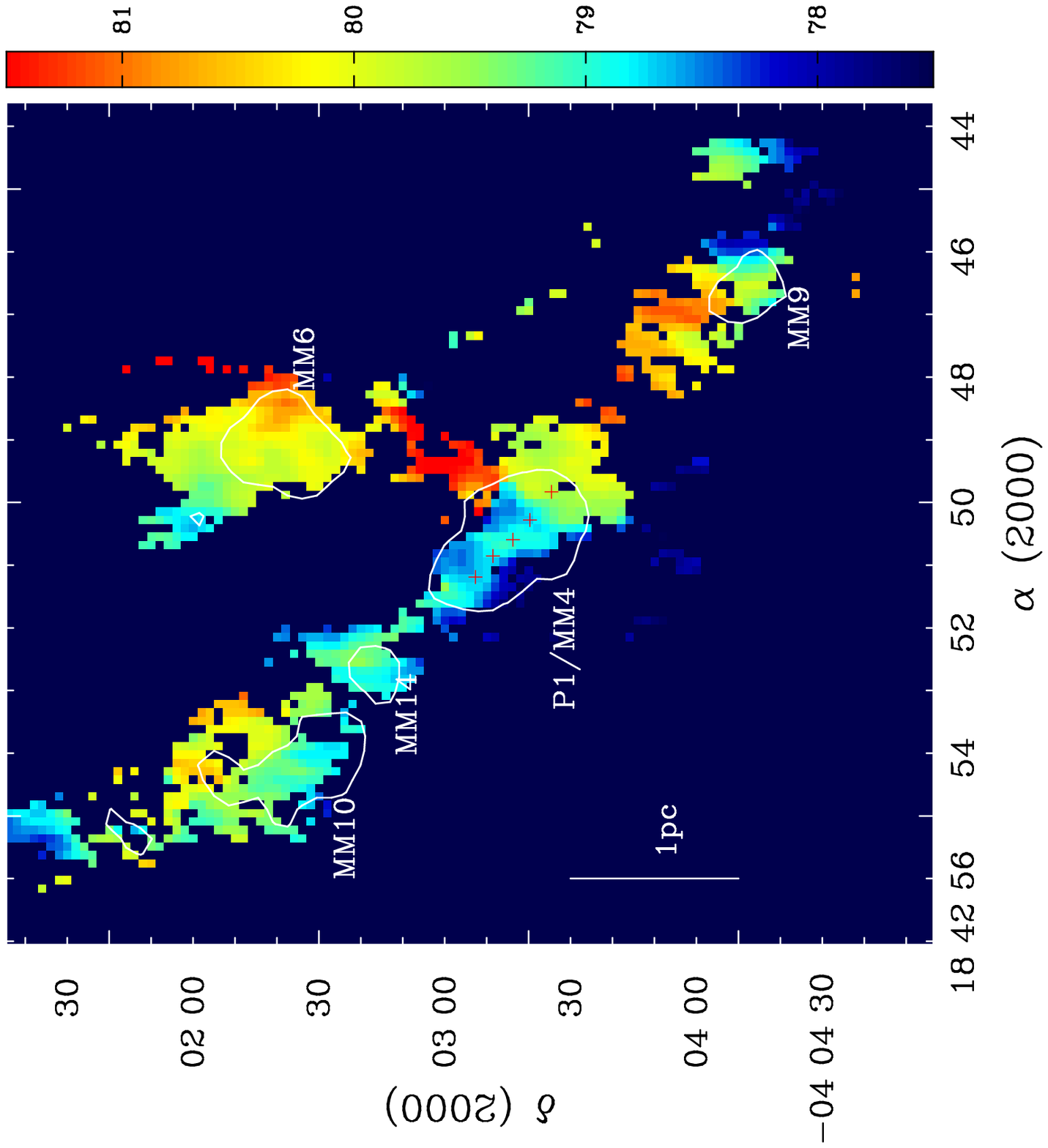}
\caption{Shown in color scales is the centroid velocity (moment 1) map in the IR dark region of G28.34 derived from the \nh3 (J,K)=(1,1) emission obtained from the VLA. The white contour outlines the 1.3mm continuum emission from the IRAM 30m Telescope \citep{rathborne2010}. The scale bar on the right-hand side of the plot denotes velocity in \kms-1. The $+$ symbol marks the five continuum peaks (SMA1 through SMA5) reported in \citet{zhang2009}.}
\end{figure}

\clearpage
\begin{figure}[h]
\figurenum{7}
\includegraphics{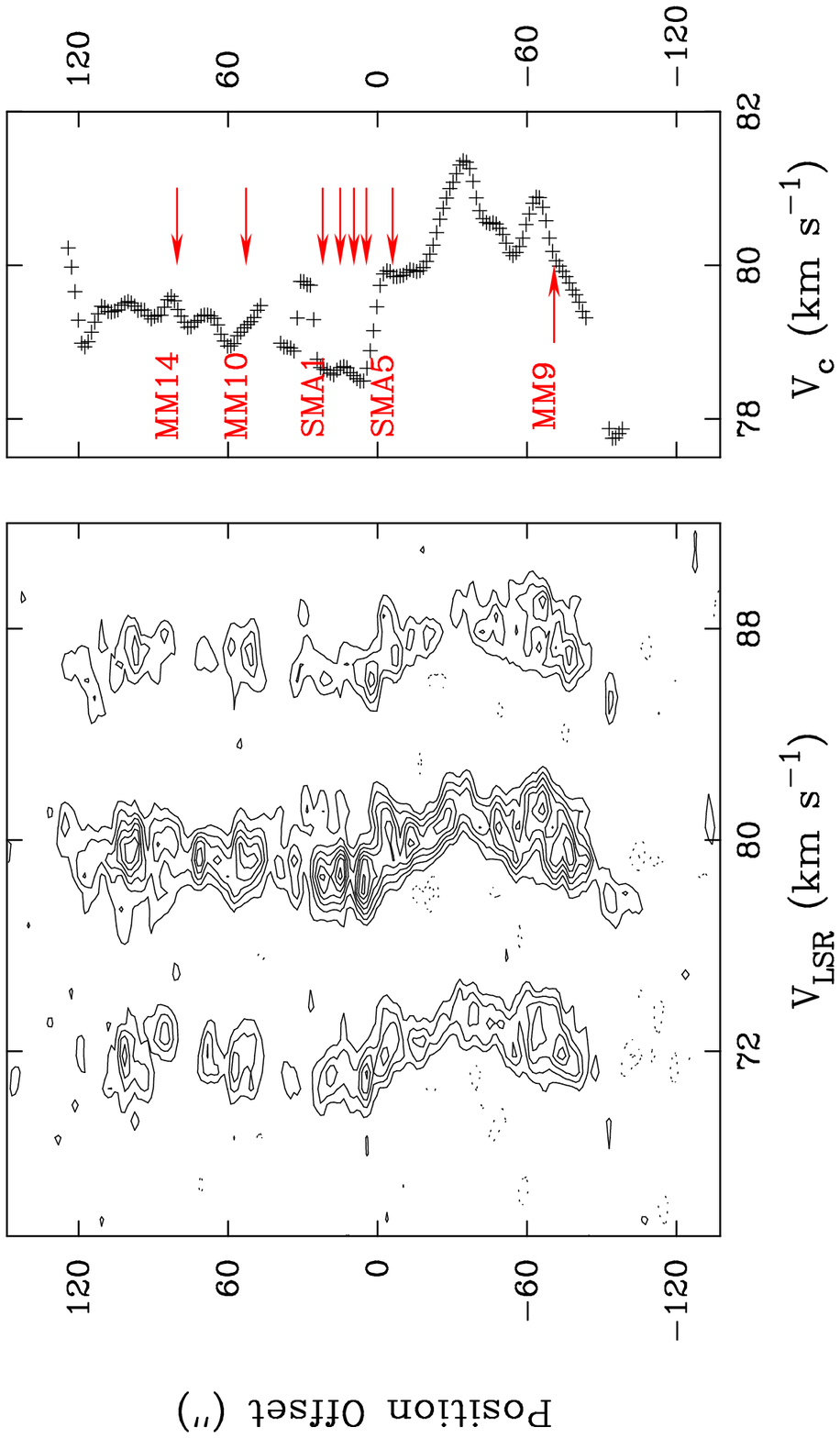}
\caption{{\bf Left:} Position velocity (PV) diagram of the \nh3 (1,1) emission. The cut is along the main filament with an offset close to Core 3. {\bf Right:} The centroid velocity along the PV cut. The arrows mark the locations of dust continuum sources. }
\end{figure}

\clearpage
\begin{figure}[h!]
\figurenum{8}
\includegraphics{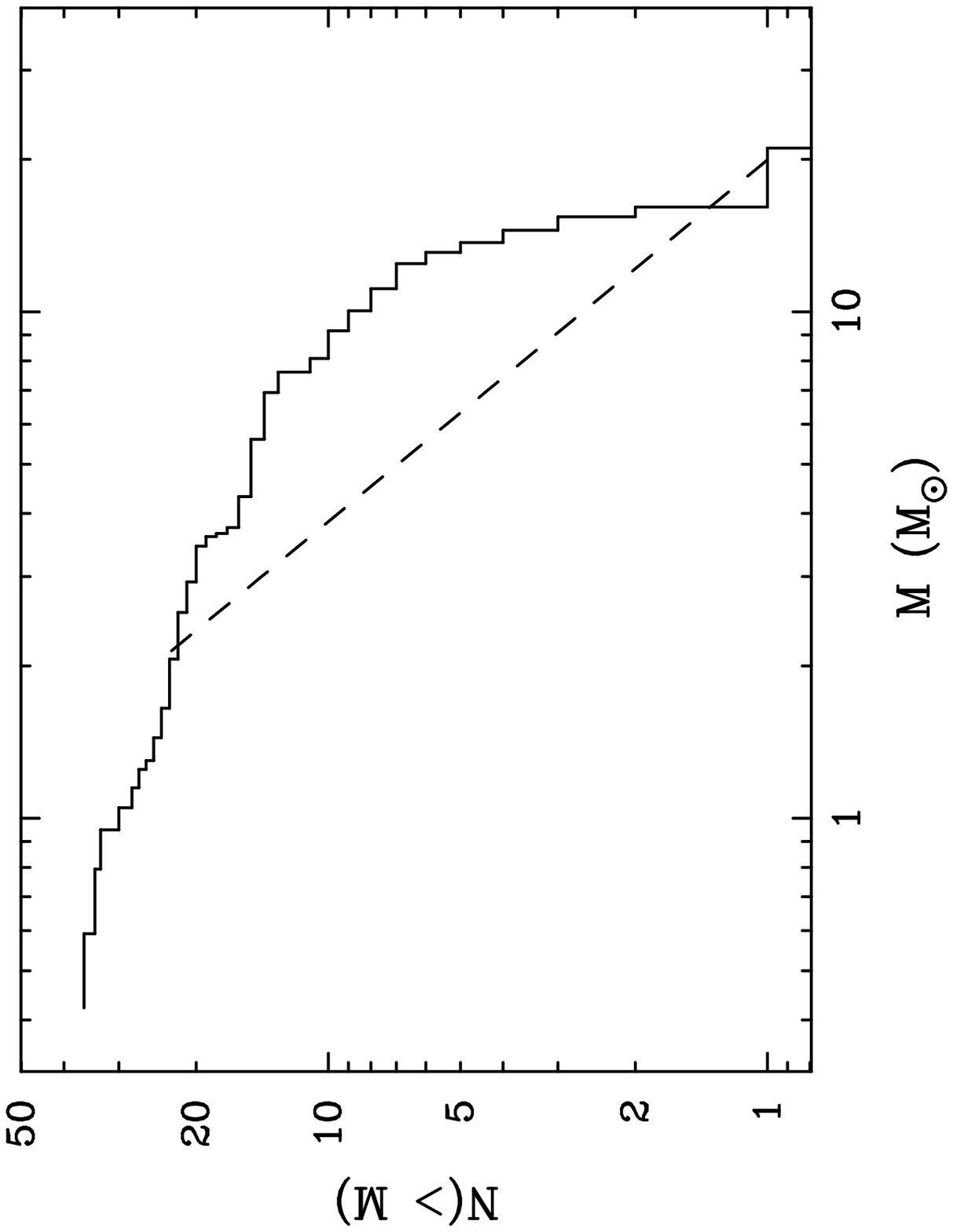}
\caption{Cumulative mass function of the fragments identified in G28.34 P1. The data are plotted in logarithmic scales. The dashed line marks the slope of the Salpeter initial mass function.}
\end{figure}

\clearpage
\begin{figure}[h!]
\figurenum{9a}
\includegraphics{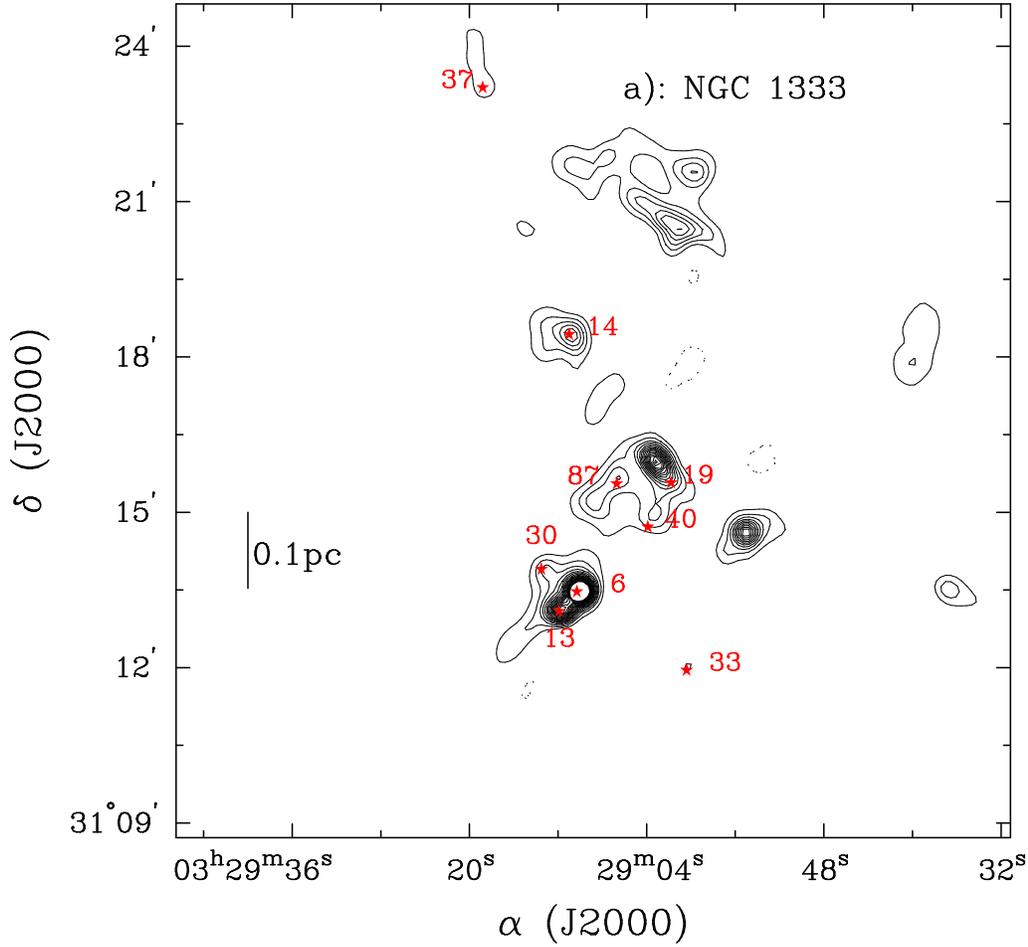}
\caption{{\bf a):} Dust continuum image at 870 $\mu$m obtained from JCMT of the protocluster NGC 1333. The star symbol marks protostars reported in \citet{sadavoy2014}. {\bf b):}  Simulated ALMA observations using the clean components of the 1.3mm continuum for G28.34. The simulation adopts the observing conditions and the on-source integration time of the ALMA project. The star symbol in panels b) and c) marks the brightest 1.3mm continuum peaks shown in Figure 2. {\bf c):}  Simulated ALMA observations made from the clean components of G28.34 and the model from the NGC 1333 protocluster. The east-west direction in the NGC 1333 model image is reversed.}
\end{figure}

\clearpage
\begin{figure}[h!]
\figurenum{9b}
\includegraphics{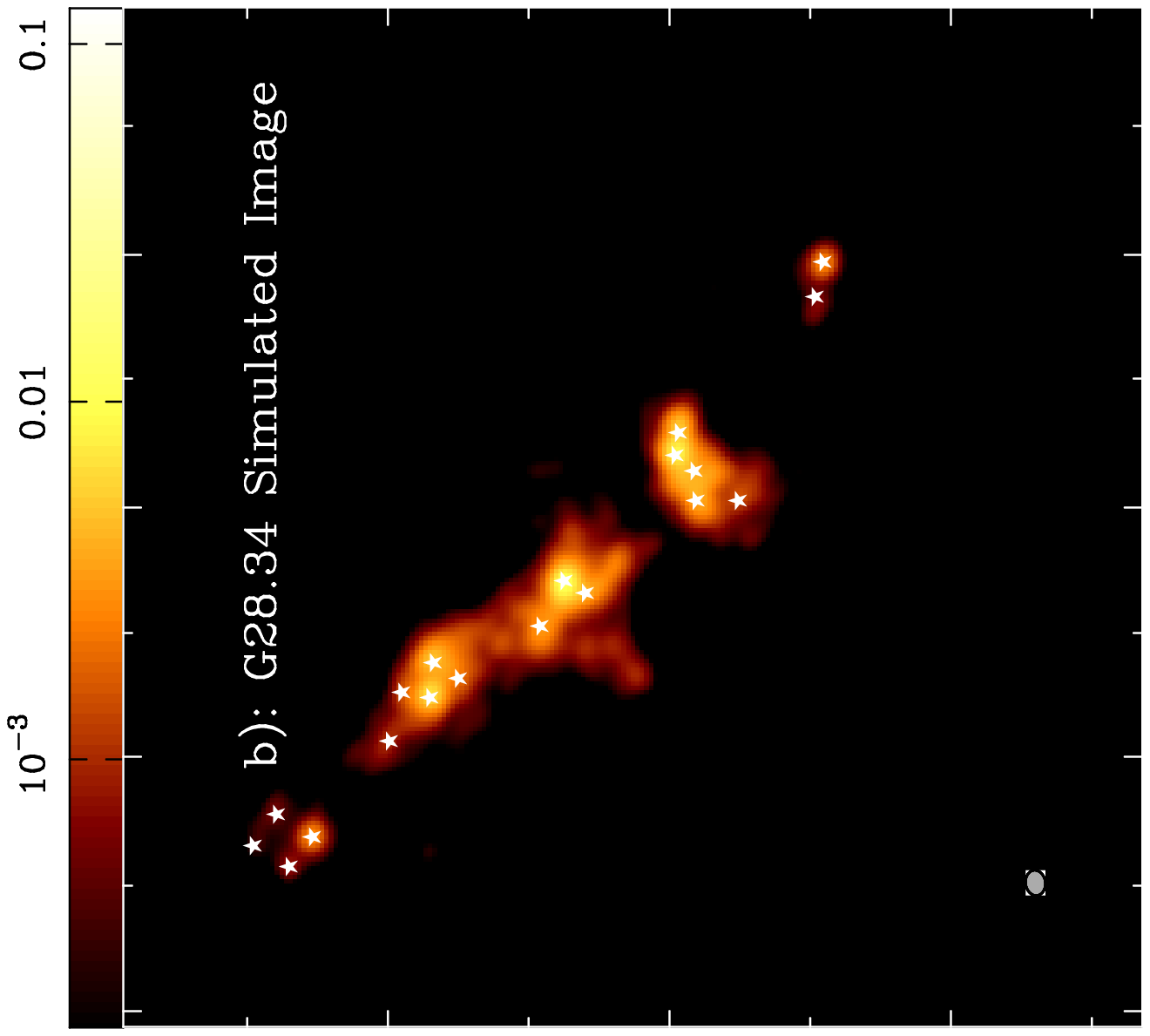}
\caption{}
\end{figure}

\clearpage
\begin{figure}[h!]
\figurenum{9c}
\includegraphics{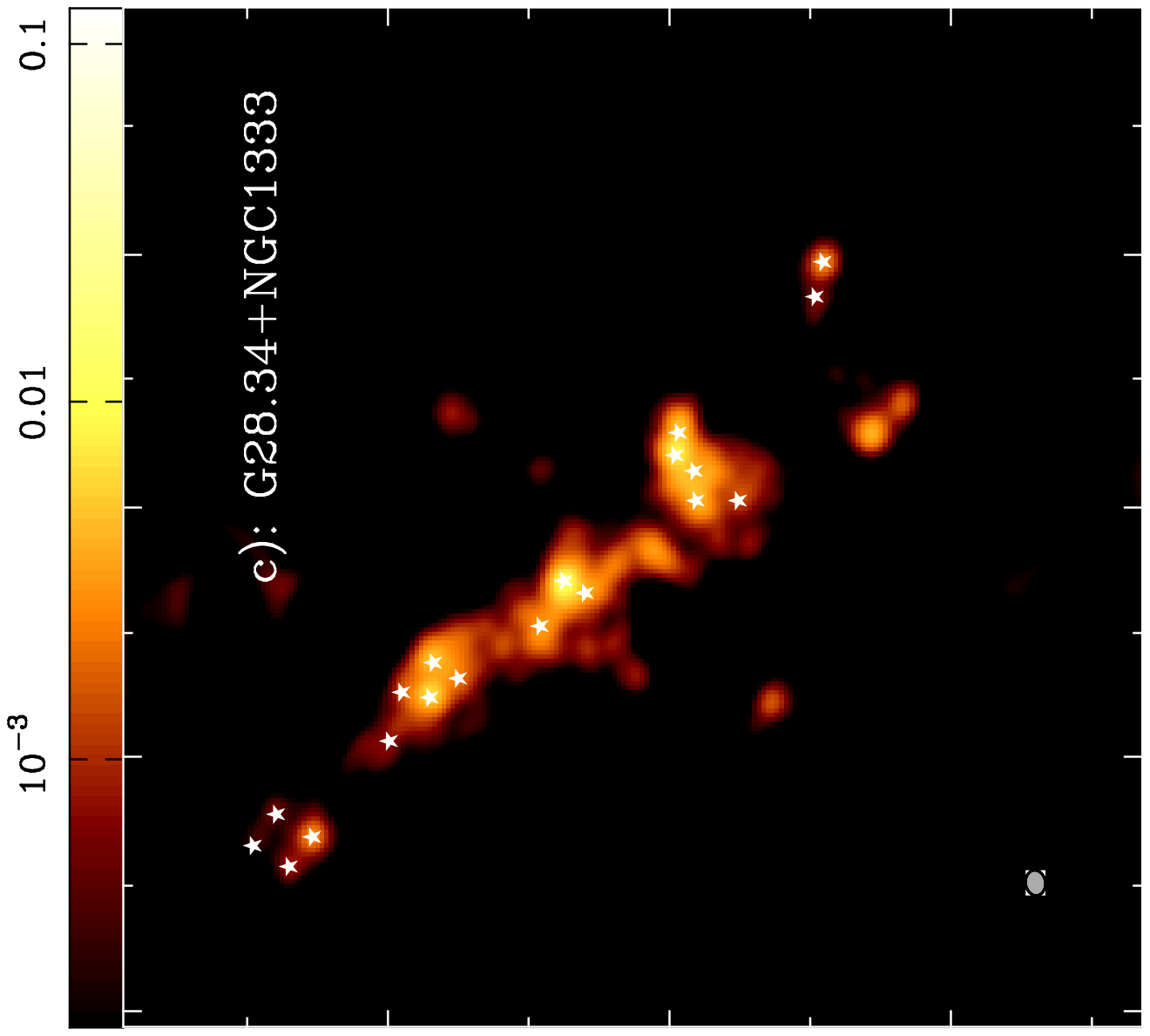}
\caption{}
\end{figure}

%\input{ms_rev.bbl}
%\input{/home/qzhang/tex/acronyms}
%\bibliography{/home/qzhang/tex/bibliography}
%\bibliographystyle{/home/qzhang/tex/aa} % this does the style, aa.bst

\end{document}